\documentclass[%
 reprint,
superscriptaddress,
 amsmath,amssymb,
 aps,
prb,
]{revtex4-1}
\usepackage{graphicx}
\usepackage{amsmath}
\usepackage{placeins}
\usepackage{color}
\usepackage{bm}
\newcommand{\SSS}{\mathbf{S}}
\newcommand{\qqq}{\mathbf{q}}
\newcommand{\QQQ}{\mathbf{Q}}
\newcommand{\aaa}{\mathbf{a}}
\newcommand{\bbb}{\mathbf{b}}
\newcommand{\ccc}{\mathbf{c}}

\newcommand{\vvv}{\mathbf{v}}

\newcount\colveccount
\newcommand*\colvec[1]{
        \global\colveccount#1
        \begin{pmatrix}
        \colvecnext
}
\def\colvecnext#1{
        #1
        \global\advance\colveccount-1
        \ifnum\colveccount>0
                \\
                \expandafter\colvecnext
        \else
                \end{pmatrix}
        \fi
}

\begin{document}
\title{Spin dynamics and exchange interactions in CuO measured by neutron scattering}
\author{H. Jacobsen}
\affiliation{Department of Physics, Oxford University, Clarendon Laboratory, Oxford, OX1 3PU, United Kingdom}
\author{S. M. Gaw }
\affiliation{Department of Physics, Oxford University, Clarendon Laboratory, Oxford, OX1 3PU, United Kingdom}
\author{A. J. Princep}
\affiliation{Department of Physics, Oxford University, Clarendon Laboratory, Oxford, OX1 3PU, United Kingdom}
\author{ E. Hamilton}
\affiliation{Department of Physics, Oxford University, Clarendon Laboratory, Oxford, OX1 3PU, United Kingdom}

\author{S. T\'oth}
\affiliation{Laboratory for Neutron Scattering and Imaging, Paul Scherrer Institut, CH-5232 Villigen, Switzerland}

\author{R. A. Ewings}
\affiliation{ISIS Pulsed Neutron and Muon Source, STFC Rutherford Appleton Laboratory, Harwell Campus, Didcot, OX11 0QX, United Kingdom}

\author{M. Enderle}
\affiliation{Institut Laue--Langevin, 71 Avenue des Martyrs, CS 20156, 38042 Grenoble Cedex 9, France}

\author{ E. M. H\'etroy Wheeler}
\affiliation{Institut Laue--Langevin, 71 Avenue des Martyrs, CS 20156, 38042 Grenoble Cedex 9, France}

\author{ D. Prabhakaran}
\affiliation{Department of Physics, Oxford University, Clarendon Laboratory, Oxford, OX1 3PU, United Kingdom}

\author{ A. T. Boothroyd }
\affiliation{Department of Physics, Oxford University, Clarendon Laboratory, Oxford, OX1 3PU, United Kingdom}

\date{\today}

\begin{abstract}
The magnetic properties of CuO encompass several contemporary themes in condensed matter physics, including quantum magnetism, magnetic frustration, magnetically-induced ferroelectricity and orbital currents. Here we report polarized and unpolarized neutron inelastic scattering measurements which provide a comprehensive map of the cooperative spin dynamics in the low temperature antiferromagnetic (AFM) phase of CuO throughout much of the Brillouin zone. At high energies ($E \gtrsim 100$\,meV) the spectrum displays continuum features consistent with the des Cloizeax--Pearson dispersion for an ideal $S=\frac{1}{2}$ Heisenberg AFM chain. At lower energies the spectrum becomes more three-dimensional, and we find that a linear spin-wave model for a Heisenberg AFM provides a very good description of the data, allowing for an accurate determination of the relevant exchange constants in an effective spin Hamiltonian for CuO. In the high temperature helicoidal phase, there are features in the measured low-energy spectrum that we could not reproduce with a spin-only model. We discuss how these might be associated with the magnetically-induced multiferroic behavior observed in this phase.

\end{abstract}
\maketitle
\section{Introduction}

Despite its simple chemical formula, cupric oxide (CuO, tenorite) has a surprisingly complex magnetic behavior which remains only partly understood. Its properties were investigated extensively following the discovery of the copper oxide high-temperature superconductors \cite{Bednorz1986}, when it was hoped that insights from studying CuO might lead to an improved understanding of the mechanism of superconductivity. However, despite some similarities in structure and bonding the essential physics of CuO is fundamentally different from that of the copper oxide superconductors. The key ingredient of the latter is the charge-doped CuO$_2$ layers which host quasi-two-dimensional antiferromagnetic (AFM) spin correlations \cite{Keimer2015}, whereas CuO is an insulator which has not so far been made metallic by doping and whose magnetic behavior is dominated by quasi-one-dimensional (1D) AFM chains \cite{OKeeffe1962,Forsyth1988,Yang1988,Yang1989,Ain1989,Boothroyd1997,Shimizu2003}.

The discovery of improper ferroelectricity at temperatures between 213\,K and 230\,K associated with a helicoidal magnetic structure  has renewed interest in CuO \cite{Kimura2008}, identifying it as a model system for the study of magnetically-induced multiferroicity and related  magnetoelectric and orbital phenomena \cite{Scagnoli2011,Babkevich2012,Johnson2012,Jones2014,Cao2015a}. The stability of this phase at relatively high temperatures has raised the prospect that multiferroic behaviour at room temperature might be achievable in CuO under pressure \cite{Rocquefelte2012,Rocquefelte2013}, or through doping \cite{Hellsvik2014}. Theoretical studies have identified strong superexchange interactions along the AFM chains and frustrated inter-chain interactions as key ingredients in the multiferroic phase, but the detailed mechanism remains under debate \cite{Toledano2011a,Giovannetti2011a,Jin2012,Villarreal2012,Apostolova2013}.

Progress in developing and validating models for the complex magnetic and magnetoelectric behavior of CuO has been hampered by the lack of reliable information on the exchange interactions. Previous neutron and optical spectroscopic measurements have probed parts of the spin excitation spectrum and reported approximate values for a few nearest-neighbour exchange constants \cite{Yang1989,Ain1989,Chattopadhyay1992,Boothroyd1997,Jung2009,Choi2013}. Larger sets of exchange interactions have been calculated by \textit{ab initio} methods \cite{Filippetti2005,Rocquefelte2010,Giovannetti2011a,Jin2012,Pradipto2012}. These studies all agree on a dominant AFM interaction along the $[10\bar{1}]$ direction, consistent with the observed quasi-one-dimensional magnetic behavior, but there is no consensus on which of the other exchange constants are relevant or on their relative strengths.  An overview of the exchange parameters as determined in some of the recent experimental and theoretical studies of CuO is given in Table~\ref{tab:model_parameters}.

The aim of this paper is to advance the understanding of CuO through the development of a spin Hamiltonian which can be used as the basis for theoretical models.  To this end, we have measured the momentum-resolved magnetic spectrum of antiferromagnetically-ordered CuO throughout much of the Brillouin zone by inelastic neutron scattering. We find that the form of the magnetic spectrum crosses over from quasi-one-dimensional at high energies to three-dimensional below about 100\,meV, and we show that the inter-chain coupling is strongly frustrated. Using linear spin-wave theory to model the spectrum we have identified the relevant exchange interactions and obtained experimental values for them. We find substantial differences between some of the exchange parameters so-obtained and those reported previously.

The paper is organized as follows. In the next section we review details of the crystal and magnetic structure of CuO, and then we present our neutron scattering results and analysis in terms of a spin Hamiltonian, starting with the low temperature AFM phase before moving on to the helicoidal phase. We end with some discussion of the significance of our findings in relation to the properties of CuO.

\begin{table*}
\caption{Exchange parameters for CuO. The corresponding exchange paths are shown in Fig.~\ref{fig:exchange_interactions}. The parameters $D_1$ and $D_2$ are exchange anisotropy parameters used in this work. All values are given in meV. The numbers in parentheses are estimated errors in the last digits, obtained either from standard fitting routines or by varying the parameters until a noticable worsening in the quality of agreement between model and data occurred. Some of the parameters, and hence their uncertainties, are correlated. 
}
{%
\newcommand{\mc}[3]{\multicolumn{#1}{#2}{#3}}
\begin{center}
\begin{tabular}{c|ccccccccccc}
\hline \hline
\textit{Ab initio}  & $J$ & $J_{ac}$ & $J_b$ & $J_{ab}$ & $J_{bc}$ & $\delta$ & $J_{2,a}$ & $J_{2,c}$ &  $J_{2,ab}$ & $D_1$ &$D_2$\\ \hline
Ref.~\onlinecite{Filippetti2005} & $38.4$ & $-20.4$ & $-$ & $-8$ & $-11.6$ & $-$ & $14$ & $14$ & $-$ & $-$ & $-$ \\
Ref.~\onlinecite{Rocquefelte2010} ($\alpha=0.15$) & $128.8$ & $-2.6$ & $-$ & $18.2$ & $-4.2$  & $-$ & 0 & $30.1$ & 0& $-$ & $-$ \\
Ref.~\onlinecite{Rocquefelte2010} ($\alpha=0.25$) & $80.5$ & $-3.0$ & $-$ & $4.0$ & $-3.5$  & $-$ &0 & $19.6$ & 0 & $-$ & $-$ \\
Ref.~\onlinecite{Rocquefelte2011a} & $107.12$ & $-3.65$ & 0.77 & $8.32$ & $-2.92$  & $-$ &$-1.04$ & $10.11$ & 20.05 & $-$ & $-$ \\
Ref.~\onlinecite{Giovannetti2011a}  ($U_\text{eff}=5.5) \footnote{Some of the exchange constants were mislabeled in Ref.~\onlinecite{Giovannetti2011a}, as pointed out in Ref.~\onlinecite{Rocquefelte2011a} and acknowledged in Ref.~\onlinecite{Giovannetti2011b}. We give the corrected labels here.}$ & $107.76$  & $-15.76$& $-21.48$ & $15.82$ & $7.98$ & $-$ & $16.18$ & $6.89$ & $-$ & $-$ & $-$ \\
Ref.~\onlinecite{Giovannetti2011a}  ($\alpha=0.15)^\textrm{a}$ & $120.42$ & $-24.33$ & $-23.02$ & $13.17$ & $4.19$ & $-$ & $14.27$ & $4.99$ & $-$ & $-$ & $-$ \\
Ref.~\onlinecite{Jin2012} & $51$ & $-8.6$ & $-9.87$ & $-4.9$ & $-7$  & $-$& $12$ & $-2.1$ & $-$ & $-$ & $-$ \\
Ref.~\onlinecite{Pradipto2012} & $47.5  $ & $0.8$ & $-$ & $-9.0$ & $-3.7$  & $-$& $-$ & $-$  & $5.1$ & $-$ & $-$\\
Ref.~\onlinecite{Ganga2017}       & $127.48$ & $-8.6$ & $-$ & $33.18$ & $3.29$  & $-$ &  $39.11$ &$-$ & $-$ & $-$ & $-$ \\
Ref.~\onlinecite{Dai2004}      & $75.0$ & $4.7$ & $-$ & $0.24$ & $4.0$ & $-$ & $-$ & $-$ & $3.9$ & $-$ & $-$ \\
\hline \hline
 Experiments  & $J$ & $J_{ac}$ & $J_b$ & $J_{ab}$ & $J_{bc}$ & $\delta$ & $J_{2,a}$ & $J_{2,c}$  & $J_{2,ab}$ & $D_1$ &$D_2$\\ \hline
Ref.~\onlinecite{Shimizu2003} (susceptibility) & $77(3)$ & $-$ & $-$ & $-$ & $-$ & $-$ & $-$ & $-$ & $-$ & $-$ & $-$ \\
 Ref.~\onlinecite{Choi2013} (Raman) & $108$ & $-$ & $-$ & $-$ & $-$ & $-$ & $-$ & $-$ & $-$ & $-$ & $-$ \\
Ref.~\onlinecite{Jung2009} (infra-red) & $95$--$100$ & $-$ & $-$ & $-$ & $-$ & $-$ & $-$ & $-$ & $-$ & $-$ & $-$ \\
 Ref.~\onlinecite{Ain1989} (neutron)\footnote{The exchange parameters given in Ref.~\onlinecite{Ain1989} have been multiplied by a factor of 2 to account for differences in the definition of $\mathcal{H}$. In addition,   $J$  given in Ref.~\onlinecite{Ain1989} is the renormalized exchange parameter  $J^\textrm{sw}$ obtained by LSWT; we give here the bare exchange parameter $J = 2 J^\textrm{sw}/\pi$. The signs of $J_{ac}$ and $J_b$ quoted here have been inferred from the magnon dispersion presented in Ref.~\onlinecite{Ain1989}. }
 & $102$ & $-10$ & $-6$ & $-$ & $0.22$ &  $-$ & $-$ & $-$ & $-$ & $-$ & $-$  \\

 Ref.~\onlinecite{Boothroyd1997} (neutron)  & $93.6\footnote{\label{fnote:J}$J$ is determined from a fit to the two-spinon spectrum for a $S=\frac{1}{2}$ Heisenberg AFM chain.}$
 & $-$ & $-$ & $-$ & $-$ &  $-$ & $-$ & $-$ & $-$ & $-$ & $-$  \\
 This work (neutron)\footnote{ All  parameters apart from $J$ are renormalized exchange parameters obtained by LSWT with $S=\frac{1}{2}$.}   & $91.4(5)^\textrm{c}$ & $-3.73(3)$ & $-0.39(10)$ & $\pm 2.50(18)$\footnote{\label{fnote:AF2} In our model we imposed the constraints $J_{ab} = J_{ab}'$ and $J_{bc} = J_{bc}'$, see Fig.~\ref{fig:exchange_interactions}(b). $J_{ab}$ and $J_{bc}$ have been determined by mean-field theory from the propagation vector and polarised neutron scattering data in the AF2 phase. $J_{ab}$ and $J_{bc}$ have the same sign, but our model does not depend on the sign as indicated by the $\pm$ signs.} & $\pm 3.10(18)^\textrm{e}$ & $0.68(5)$\footnote{\label{fnote:Gamma}$\delta$ is defined in Eq.~(\ref{eq:delta}). The value quoted here is the low temperature value determined from the optic mode gap in the AF1 phase at $T=2$~K.} & $-$ & $-$ &  $3.17(3)$\footnote{In our model we imposed the constraint $J_{2,ab} = J_{2,ab}'$.} & $-0.015(4)$ & $0.15(2)$ \\
\hline \hline
\end{tabular}
\end{center}
}%
 \label{tab:model_parameters}
\end{table*}
    
\section{Crystal and magnetic structure of CuO}
\label{sec:structure}
\begin{figure}[ht!]
\centering
\includegraphics[width=0.42\textwidth,trim={0 0 0 1cm}]{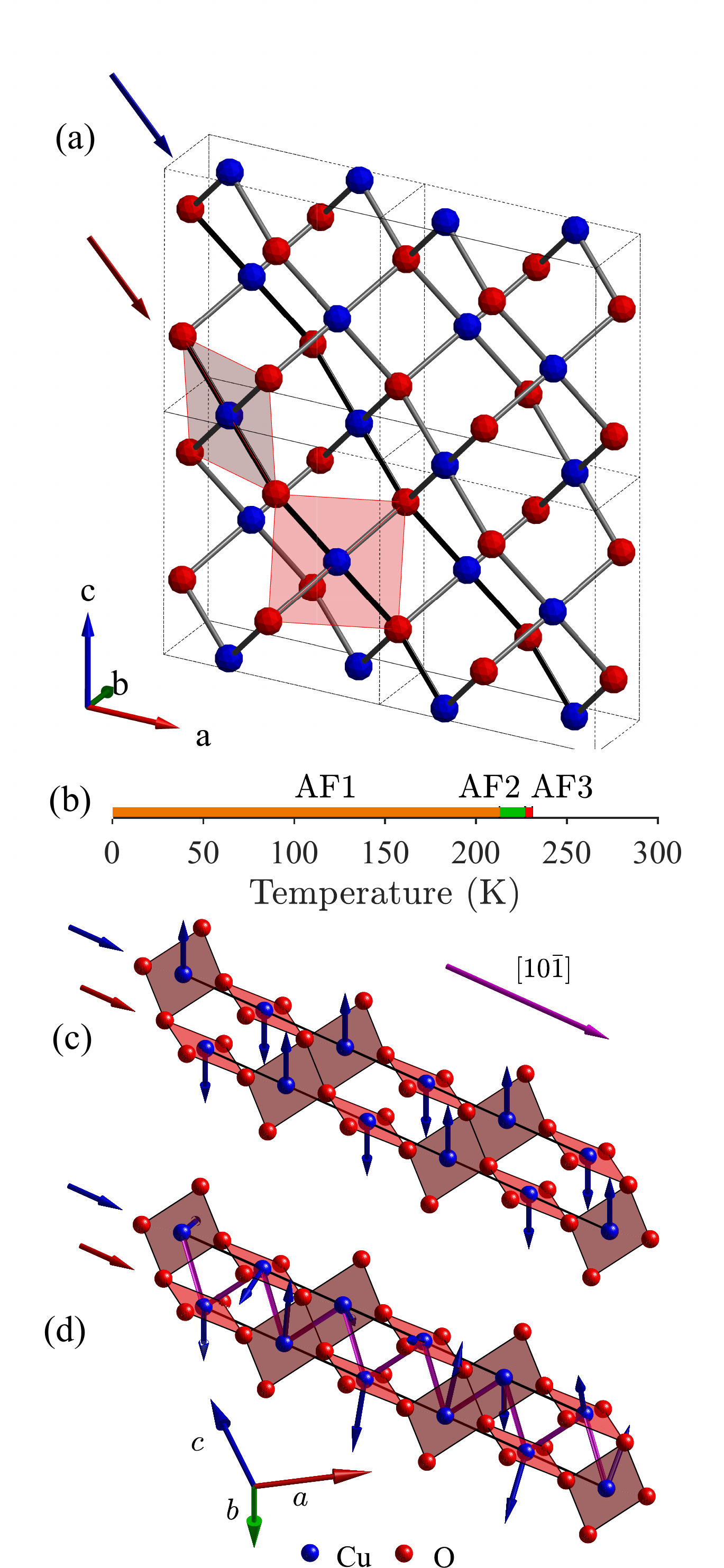}
 \caption{Crystal and magnetic structures of CuO. (a) Four unit cells of the crystal structure of CuO, with two CuO$_4$ oxygen plaquettes highlighted. The red and blue arrows indicate the two chains shown in (c) and (d). 
 (b) Schematic of the phase diagram, showing the AF1, AF2 and AF3 phases. (c) The AF1 magnetic structure on two $[10\bar{1}]$ chains. (d) The AF2 magnetic structure on two $[10\bar{1}]$ chains. The rotation angle of the helix has been exaggerated. The purple line shows how the AF2 structure can be viewed as a zig-zag chain.
}
 \label{fig:unit_cell}
\end{figure}

The crystal structure of CuO, shown in Fig.~\ref{fig:unit_cell}(a), is monoclinic with space group either $Cc$ or $C2/c$ (with the same unit cell) at room temperature\cite{Asbrink1970,Forsyth1988,Asbrink1991}.
Each Cu atom is surrounded by four  coplanar O atoms making an approximately square plaquette. Two, more distant, O atoms above and below the plaquette complete a highly distorted octahedron. 
The connected structure can be regarded as having two types of buckled Cu--O chains, running along the $[1,0,\overline{1}]$ and $[1,0,1]$ directions with Cu--O--Cu bond angles of $146^\circ$ and $109^\circ$, respectively.

The conventional unit cell is base-centred on the $ab$ face [see Fig.~\ref{fig:unit_cell}(a)] and contains four CuO molecules. The room temperature cell parameters are $a=4.684$~\AA{}, $b=3.423$~\AA{} and $c=5.129$~\AA{}, with $\beta=99.54^\circ$ (Ref.~\onlinecite{Asbrink1970}).
There are, however, indications in the literature that the room temperature structure could be distorted. In particular, weak superstructure reflections have been observed by electron diffraction\cite{Zheng2000}, and a doubling of the unit cell $\{ \aaa,\bbb,\ccc \} \rightarrow \{ \aaa+\ccc,\bbb,\aaa-\ccc \}$ was suggested to explain forbidden modes observed in infrared spectra  \cite{Kuzmenko2001}. We shall return to this point later. The crystal structure has not been refined in detail in the magnetic phases, although any centre of symmetry present in the paramagnetic phase must inevitably disappear in the ferroelectric phase.

\begin{figure}[b]
\centering
 \includegraphics[width=0.45\textwidth]{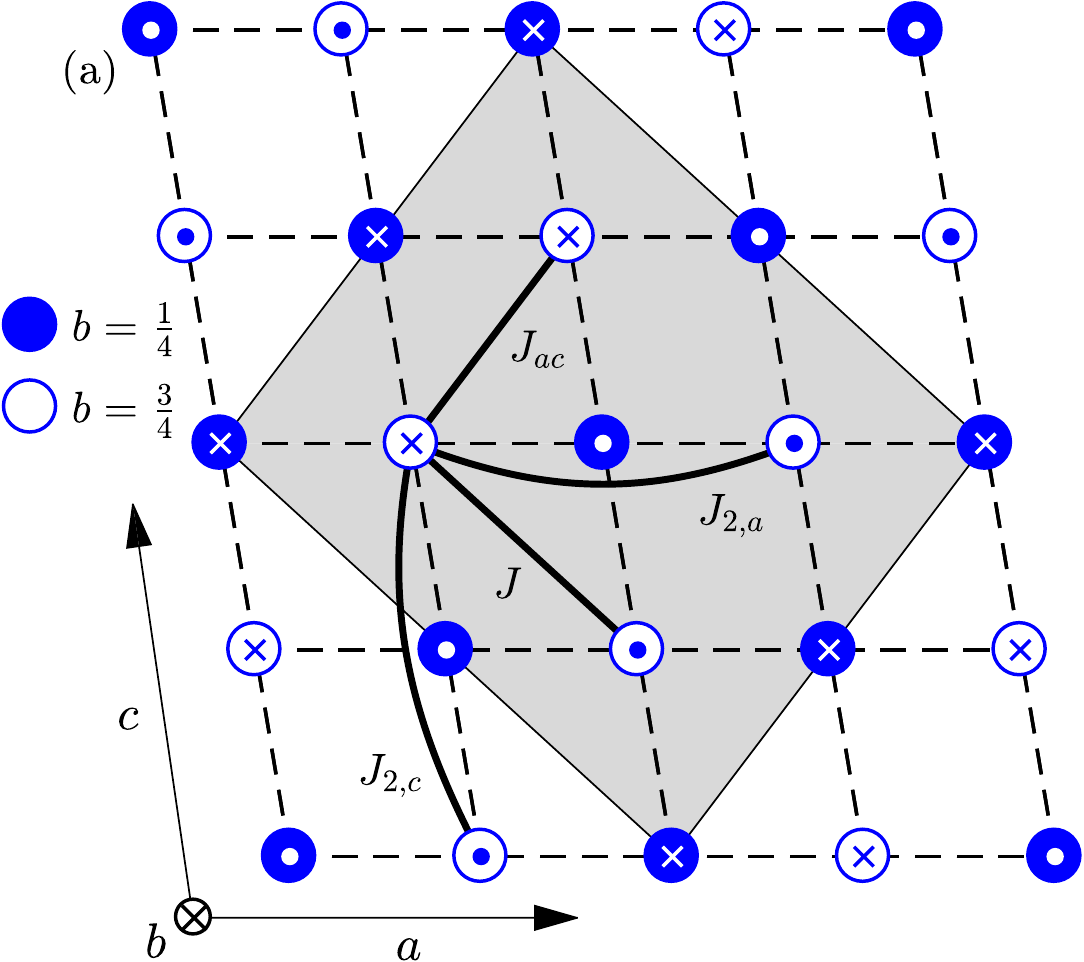} \\
 \vspace{0.5cm}
 \includegraphics[height=0.13\textheight]{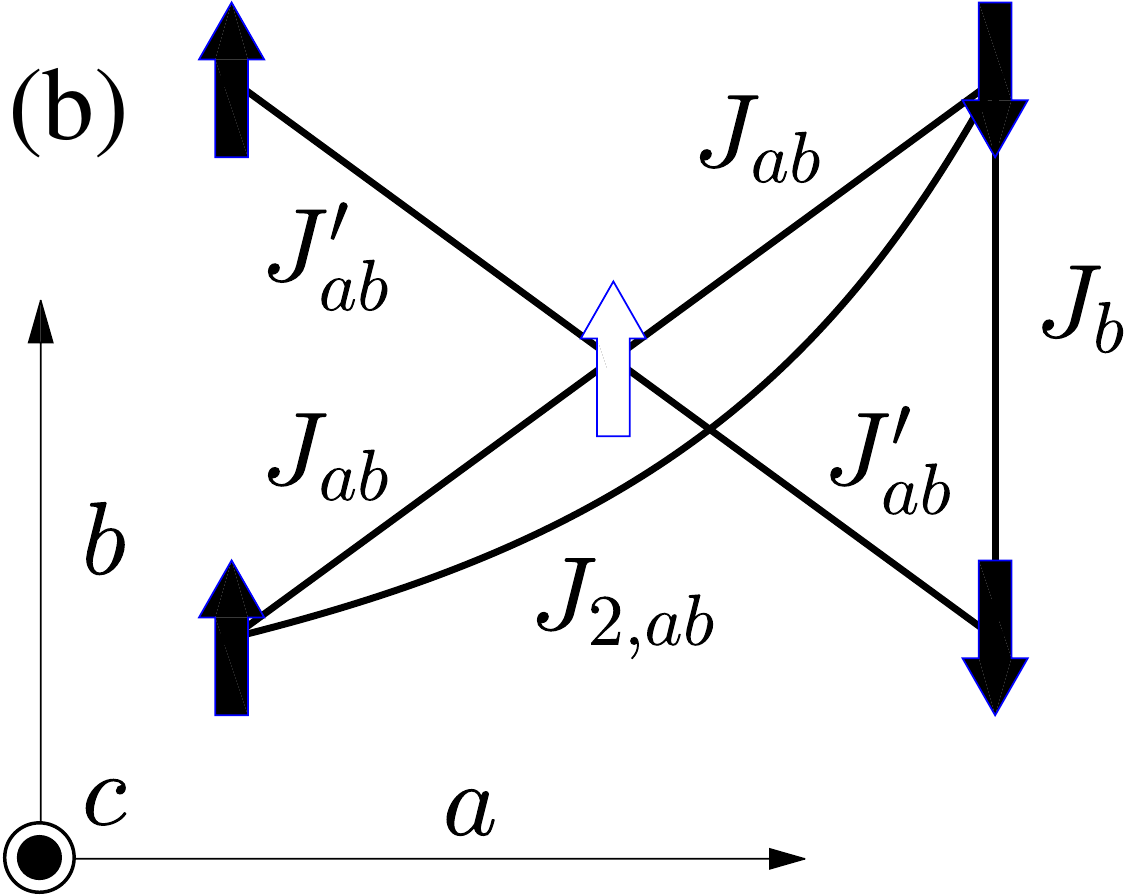}
\hfill
\includegraphics[height=0.13\textheight]{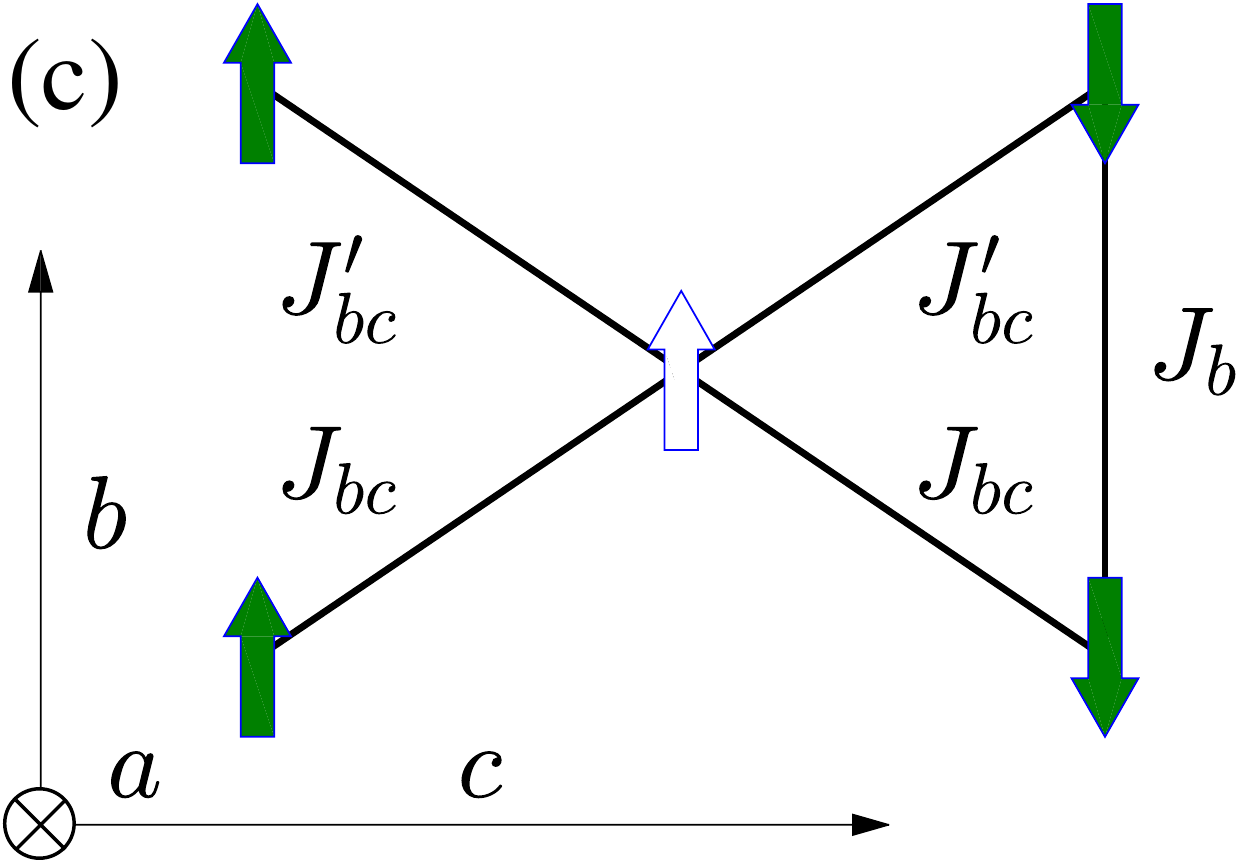}
 \caption{The AF1 magnetic structure of CuO showing the exchange interactions considered here. (a) Projection down the $b$ axis onto the $ac$ plane, with dots (crosses) used to indicate spins pointing out of (into) the plane. Cu spins on lattices $A$ and $B$ related by the C-centering are indicated by filled and empty circles, respectively. The gray area indicates the unit cell in the AF1 phase.
 (b) Projection down the $c$ axis onto the $ab$ plane. 
 (c) Same as (b) but projecting down the $a$ axis onto the $bc$ plane. In $C2/c$ the $J_{bc}$ and $J_{bc}'$ interactions are identical.
 The frustration caused by $(J_{ab},J_{ab}')$ and $(J_{bc},J_{bc}')$ is clear. 
 }
 \label{fig:exchange_interactions}
\end{figure}

The temperature phase diagram of CuO is outlined in Fig.~\ref{fig:unit_cell}(b).
At temperatures below 213~K, CuO displays commensurate antiferromagnetism with the spins aligned along the $b$ axis\cite{Forsyth1988,Yang1988,Yang1989,Brown1991a}. The low temperature ordered moment is approximately 0.68\,$\mu_\textrm{B}$ per Cu. The magnetic structure is described by the propagation vector $\qqq_1 = (\frac{1}{2},0,-\frac{1}{2})$ given in terms of the reciprocal lattice vectors $(\aaa^*,\bbb^*,\ccc^*)$. This is known as the AF1 phase and is illustrated in Fig.~\ref{fig:unit_cell}(c) and in Fig.~\ref{fig:exchange_interactions}.

In the ferroelectric AF2 phase, between 213~K and $\sim$230~K, CuO adopts an incommensurate helicoidal structure with propagation vector $\qqq_2 = (0.506,0,-0.483) = \qqq_1 + \mbox{\boldmath $ \epsilon$}$, where $\mbox{\boldmath $ \epsilon$} = (0.006,0,0.017)$. The spins rotate in the plane defined by $\bbb^*$ and $\vvv=0.506\aaa^*+1.517\ccc^*=0.286\aaa + 0.373\ccc$, as illustrated in Fig.~\ref{fig:unit_cell}(d)\cite{Forsyth1988,Brown1991,Ain1992,Babkevich2012}. In a narrow temperature range just above the AF2 phase there is evidence for an AF3 phase in which only half of the spins order \cite{Wang2016,Villarreal2012,Rebello2013}.  Various other  magnetic phases appear on application of a magnetic field \cite{Wang2016,Villarreal2012,Quirion2013}. 

The magnetic order and dynamics of CuO are usually discussed with respect to a Heisenberg effective spin Hamiltonian of the form
\begin{align}
 \mathcal{H}_\textrm{ex} = \sum_{\langle i,j\rangle} J_{ij}\SSS_i \cdot \SSS_j, \label{eq:Heisenberg_simple}
\end{align}
where the $J_{ij}$ are parameters for isotropic exchange interactions between pairs of spins $\textbf{S}_i$ and $\textbf{S}_j$, each pair being counted only once. Small additional terms that describe anisotropy will be discussed later. A number of potentially relevant superexchange interactions are illustrated in Fig.~\ref{fig:exchange_interactions} with respect to the AF1 magnetic structure. The dominant exchange interaction is $J$, which couples neighboring spins antiferromagnetically along the $[10\bar{1}]$ chain direction. The large Cu--O--Cu bond angle ($146^\circ$) in this direction is responsible for the sign and large magnitude of $J$ according to the Goodenough--Kanamori--Anderson rules.

Owing to the C-centering, the magnetic structure can be described by two interpenetrating AFM lattices, $A$ and $B$, with $B$ body-centered with respect to $A$ (and \textit{vice versa}). The spins are stacked ferromagnetically along the $b$ direction. Within each AFM lattice we find that the nearest-neighbour inter-chain interactions $J_{ac}$ and $J_b$ are both ferromagnetic.  {\it Ab initio} calculations (Table.~\ref{tab:model_parameters}) predict that the second-neighbor interactions $J_{2,a}$ and $J_{2,c}$ along the $a$ and $c$ axes, and $J_{2,ab}$ between spins connected by the vector $\textbf{a} + \textbf{b}$, are also non-negligible. 

Concerning the coupling between the AFM lattices, we note that a Cu spin on lattice $A$ has eight nearest neighbours on lattice $B$, four in the $ab$ plane and four in the $bc$ plane --- see Fig.~\ref{fig:exchange_interactions}(b) and (c). In the undistorted (room temperature) crystal structure, whether $Cc$ or $C2/c$, there should be two different parameters $J_{ab}$ and $J_{ab}'$ coupling nearest neighbours in the $ab$ plane,  Fig.~\ref{fig:exchange_interactions}(b), as the paths between Cu spins connected by the vectors $\pm(\textbf{a}+\textbf{b})/2$ and $\pm(\textbf{a}-\textbf{b})/2$ are inequivalent. In $Cc$ there should similarly be two parameters $J_{bc}$ and $J_{bc}'$, Fig.~\ref{fig:exchange_interactions}(c), but in  $C2/c$ these parameters are equal. In both $Cc$ and $C2/c$ the AFM lattices $A$ and $B$ are fully frustrated with respect to one another for isotropic Heisenberg couplings. In this work, as done elsewhere, we  simplify these interactions by setting $J_{ab}' = J_{ab}$ and $J_{bc}' = J_{bc}$ because our measurements cannot resolve the differences between the respective couplings. Therefore, the parameter we call $J_{ab}$ should be interpreted as the average of $J_{ab}$ and $J_{ab}'$. Similarly, our $J_{2,ab}$ parameter, see Fig.~\ref{fig:exchange_interactions}(b), should be regarded as the average of the exchanges on different diagonals $J_{2,ab}$ and $J_{2,ab}'$ which, due to experimental limitations, we cannot resolve either. 

For the sake of clarity, we give in Table~\ref{tab:exchange_paths} the vectors between the Cu spins corresponding to each exchange interaction, together with the corresponding Cu--O--Cu bond angle where it is well defined.
\begin{table*}[t]
\begin{center}
\caption{Vectors describing the different exchange paths in CuO, together with the corresponding Cu-Cu distance and Cu-O-Cu bond angle where appropriate.}
\begin{tabular}{c|ccccccccccc}
\hline \hline
 & $J$ & $J_{ac}$ & $J_b$ & $J_{ab}$ & $J'_{ab}$ & $J_{bc}$ & $J_{bc}'$ & $J_{2,a}$  & $J_{2,c}$ &  $J_{2,ab}$ &  $J_{2,ab}'$\\ \hline
Vector  & $(\frac{1}{2},0,-\frac{1}{2})$ & $(\frac{1}{2},0,\frac{1}{2})$ & $(0,1,0)$ & $(\frac{1}{2},\frac{1}{2},0)$ & $(\frac{1}{2},-\frac{1}{2},0)$ & $(0,-\frac{1}{2},\frac{1}{2})$ & $(0, \frac{1}{2},\frac{1}{2})$ & $(1,0,0)$ & $(0,0,1)$  & $(1,1,0)$ & $(1,-1,0)$\\
 Cu-Cu distance ({\AA})  & $ 3.75$  & $3.17$  & $3.42$ & $2.90$ & $2.90$ & $3.08$ & $3.08$ & $4.68$ & $5.13$ & $5.80$ & $5.80$ \\
 Cu-O-Cu angle ($^\circ$) & 146.3 & 108.4 & $-$ & 66 & 96 & 103.9 & 103.9 &$-$ &$-$ &$-$ &$-$\\
\hline
\hline
\end{tabular}%
\end{center}
\label{tab:exchange_paths}
\end{table*}

\section{Experimental details}

Single crystals of CuO were grown by the floating-zone technique \cite{Prabhakaran2003}, and characterized by susceptibility measurements which where in agreement with literature results\cite{Zheng2001,Kimura2008,Babkevich2012}. X-ray and neutron Laue diffraction were used to select crystals of high crystalline quality and to orient them.

Time-of-flight neutron scattering experiments\cite{MERLIN_data,MAPS_data} were performed on the Merlin \cite{Bewley2006} and MAPS\cite{MAPS2004} spectrometers at the ISIS Facility. For these experiments, six single crystals with a total mass of 32.5~g and individual mosaicity of $\sim 2^\circ$ were co-aligned with a resulting mosaic spread of $\sim $$3^\circ$.  The sample was mounted in a closed-cycle refrigerator, and data were recorded at a temperature of approximately 6\,K.  Multi-angle scans, in which the sample was rotated around the $b$ axis in $1^\circ$ steps, were performed with incident energies $E_\textrm{i}$ of 90, 135 and 180~meV (Merlin) and 160~meV (MAPS).  In addition, data were recorded on MAPS with $E_\textrm{i} = 300$ and 500~meV in a fixed sample orientation with the $[1, 0,-1]$ chain direction perpendicular to the incident beam  and the $b$ axis vertical. The full width at half maximum (FWHM) of the energy resolution of all these measurements is approximately 5\% of $E_\textrm{i}$ at $E = 0$, decreasing with increasing energy transfer. A standard vanadium sample was measured to allow the detector efficiencies to be normalised and all intensities to be expressed in absolute units.

Polarised inelastic neutron scattering measurements\cite{IN20_data} were performed on an individual CuO single crystal of mass 6.7~g on the IN20 spectrometer at the Institut Laue--Langevin.  A double-focusing monochromator and analyser, both made from Heusler $(111)$ crystals, were used to perform uniaxial polarisation analysis, and
a Helmholtz coil was used to change the orientation of the neutron polarisation adiabatically. Energy scans at constant scattering vector $\textbf{Q}$ were measured with a fixed final wave vector $k_\textrm{f}$ of either 2.662 or 4.1~\AA{}$^{-1}$. A pyrolytic graphite filter was placed in the scattered beam to suppress higher order neutrons. The sample was aligned with the $a$ and $c$ axes in the horizontal scattering plane and mounted in a helium cryostat to reach temperatures down to 2~K. Measurements were performed with the polarisation along the $x$, $y$ and $z$ directions of the standard Blume--Maleev coordinate system, in which $x$ is parallel to $\QQQ$, $z$ is vertical (parallel to $b$), and $y$ completes a right-handed coordinate system. 
Standard methods \cite{Blume1963,Moon1969,Maleev1961} were subsequently used to separate the magnetic signal, as outlined in Appendix~\ref{sec:neutron_detail}, and the intensities were corrected for the measured magnetic form factor of Cu by use of data in Ref.~\onlinecite{Forsyth1988}.

\section{results}
\label{sec:results}

\subsection{Intra-chain spin dynamics in the AF1 phase}

\begin{figure}
\centering
 \includegraphics[width=0.48\textwidth]{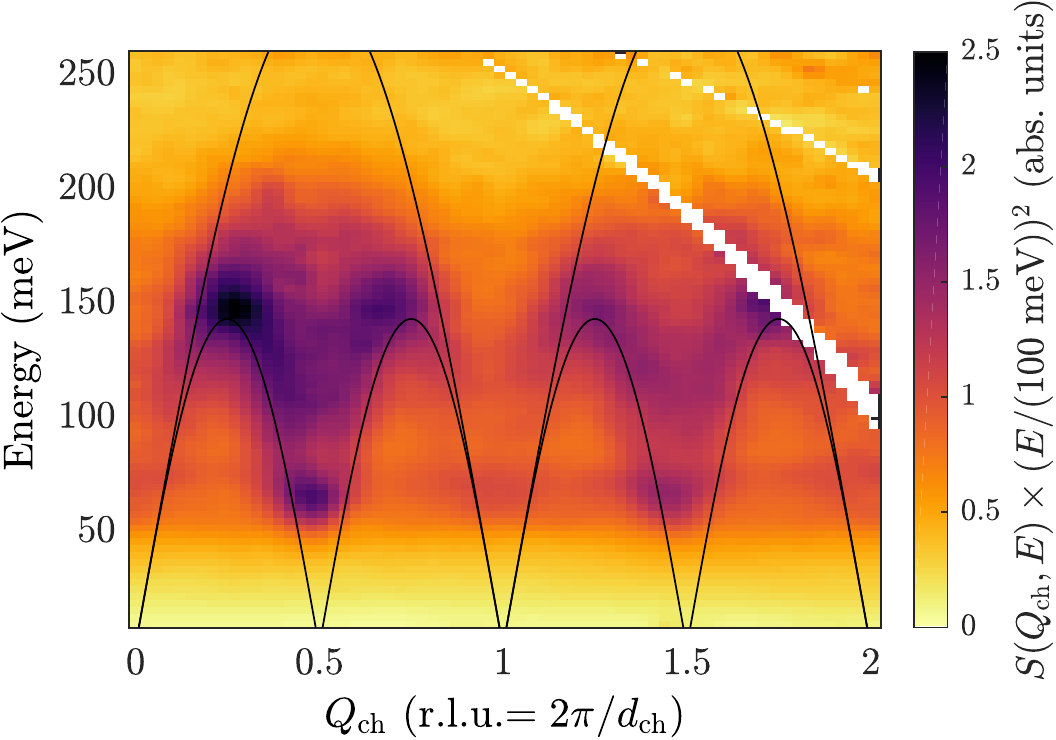}
 \caption{The spinon excitation spectrum measured in CuO by neutron scattering, showing $S(Q_\text{ch},E)$  as a function of energy transfer and wave vector along the chains, $Q_\textrm{ch}$. The spectrum has been averaged over wave vectors perpendicular to the chain direction, and an energy dependent scale factor $(E/( 100\text{ meV}))^2$  has been applied to enhance the weaker features at high energies.  The solid lines show the boundaries of the two-spinon continuum calculated from Eqs.~\eqref{eq:1d_dispa} and \eqref{eq:1d_dispb} with $J=91.4$~meV. }
 \label{fig:neutron_data_1d_model}
\end{figure}

We first consider the spin dynamics along the AFM chain direction. Fig.~\ref{fig:neutron_data_1d_model} shows the high-energy excitation spectrum measured on the MAPS spectrometer, displaying intensity as function of the scattering vector component $Q_\textrm{ch}$  along the (real space) chain direction, $[1,0,\bar{1}]$, which corresponds to the reciprocal space direction $(0.93,0,-1.09)$. A clear sinusoidal dispersion of the intensity is observed between $\sim$50~meV and $\sim$150~meV, and weak diffuse scattering is present at higher energies up to $\sim$250~meV.

We find that the high-energy scattering is consistent with the spinon spectrum of a $S=\frac{1}{2}$ Heisenberg AFM chain,  as found in a previous study of CuO (Ref.~\onlinecite{Boothroyd1997}) and in neutron scattering measurements on other Cu compounds containing quasi-1D AFM spin chains \cite{Lake2013,Zaliznyak2004,Mourigal2013}. To highlight this, we have plotted on Fig.~\ref{fig:neutron_data_1d_model} the des Cloizeaux--Pearson dispersion for an ideal $S=\frac{1}{2}$ AFM chain \cite{DesCloizeaux1962} described by the Heisenberg Hamiltonian, Eq.~\eqref{eq:Heisenberg_simple},
according to which the lower ($E_\textrm{L}$) and upper ($E_\textrm{U}$) bounds of the two-spinon continuum are given by
\begin{align}
 E_\textrm{L}(Q_\textrm{ch}) &= \frac{\pi J}{2} \left| \sin(2 \pi Q_\textrm{ch} ) \right|, \label{eq:1d_dispa}\\
 E_\textrm{U}(Q_\textrm{ch}) &= \pi J \left| \sin(\pi Q_\textrm{ch} ) \right|, \label{eq:1d_dispb}
\end{align}
where $Q_\textrm{ch}$ is in units of $2\pi/d_\textrm{ch}$, $d_\textrm{ch}$ being the separation of the spins along the chain ($d_\textrm{ch} = 3.75$~\AA{} in CuO).

In order to analyse the high-energy spectrum more quantitatively, constant-energy cuts were taken through the data  between energies of 70~meV and 250~meV (see Appendix~\ref{sec:neutron_exp_data}) and for simplicity fitted to the M\"uller Ansatz \cite{Muller1981},
\begin{align}
S_\textrm{MA}(Q_\textrm{ch},E)= \frac{A}{2\pi}\frac{\Theta(E-E_\textrm{L}(Q_\textrm{ch}))\Theta(E_\textrm{U}(Q_\textrm{ch})-E)}{ [E^2-E^2_\textrm{L}(Q_\textrm{ch})]^{1/2}},
\label{eq:MA}
\end{align}
which is an approximation to the exact two-spinon dynamical structure factor for a $S=\frac{1}{2}$ Heisenberg chain \cite{Caux2006}. In Eq.~(\ref{eq:MA}), $\Theta$ is the Heaviside step function, and $A\simeq 580$~mb\,sr$^{-1}$\,Cu$^{-1}$ is expected for an ideal $S=\frac{1}{2}$ AFM chain\cite{Lake2013}. The function $S_\textrm{MA}(Q_\textrm{ch},E)$ was convolved with the spectrometer resolution and fitted to all the cuts simultaneously by a least-squares algorithm. Best agreement was obtained with $J = 91.4 (5)$~meV and $A = 458(1)$~mb\,sr$^{-1}$\,Cu$^{-1}$ (see Appendix~\ref{sec:neutron_exp_data} for details).

The value of $J$ found here is consistent with results  from various experimental techniques (Table~\ref{tab:model_parameters}), including  the value of 93.6~meV found in a previous neutron scattering study\cite{Boothroyd1997}. The experimental value for $A$ is about 20~\% below the theoretical value, which could be due to attenuation of the neutron beam in the crystal. The good overall agreement between the data and the spinon spectrum indicates that the interaction along the chains is dominant, and that the high energy magnetic dynamics of CuO are quasi-one-dimensional.

\subsection{Inter-chain spin dynamics in the AF1 phase}
\begin{figure}
\centering
 \includegraphics[width=0.48\textwidth]{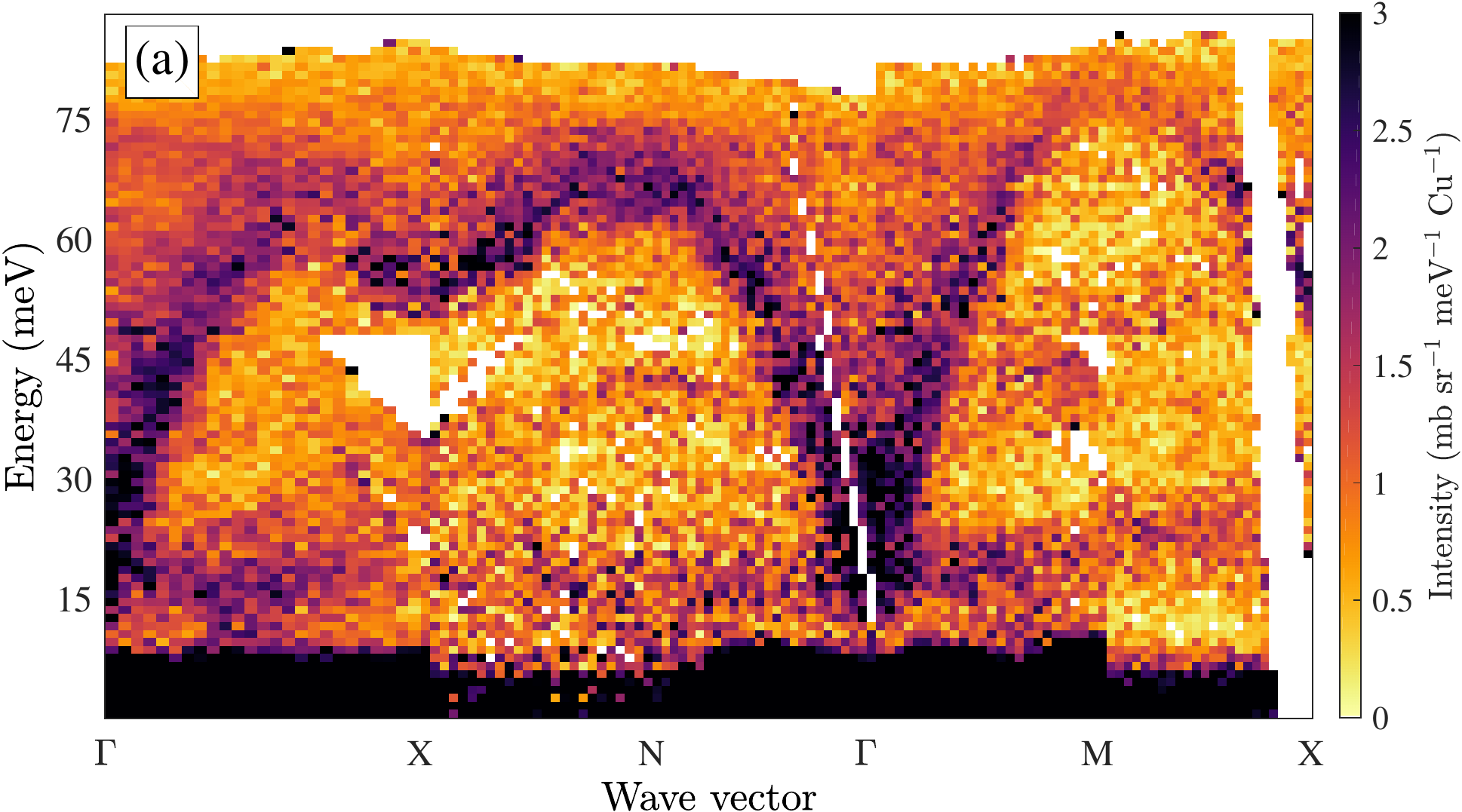} \\
  \includegraphics[width=0.48\textwidth]{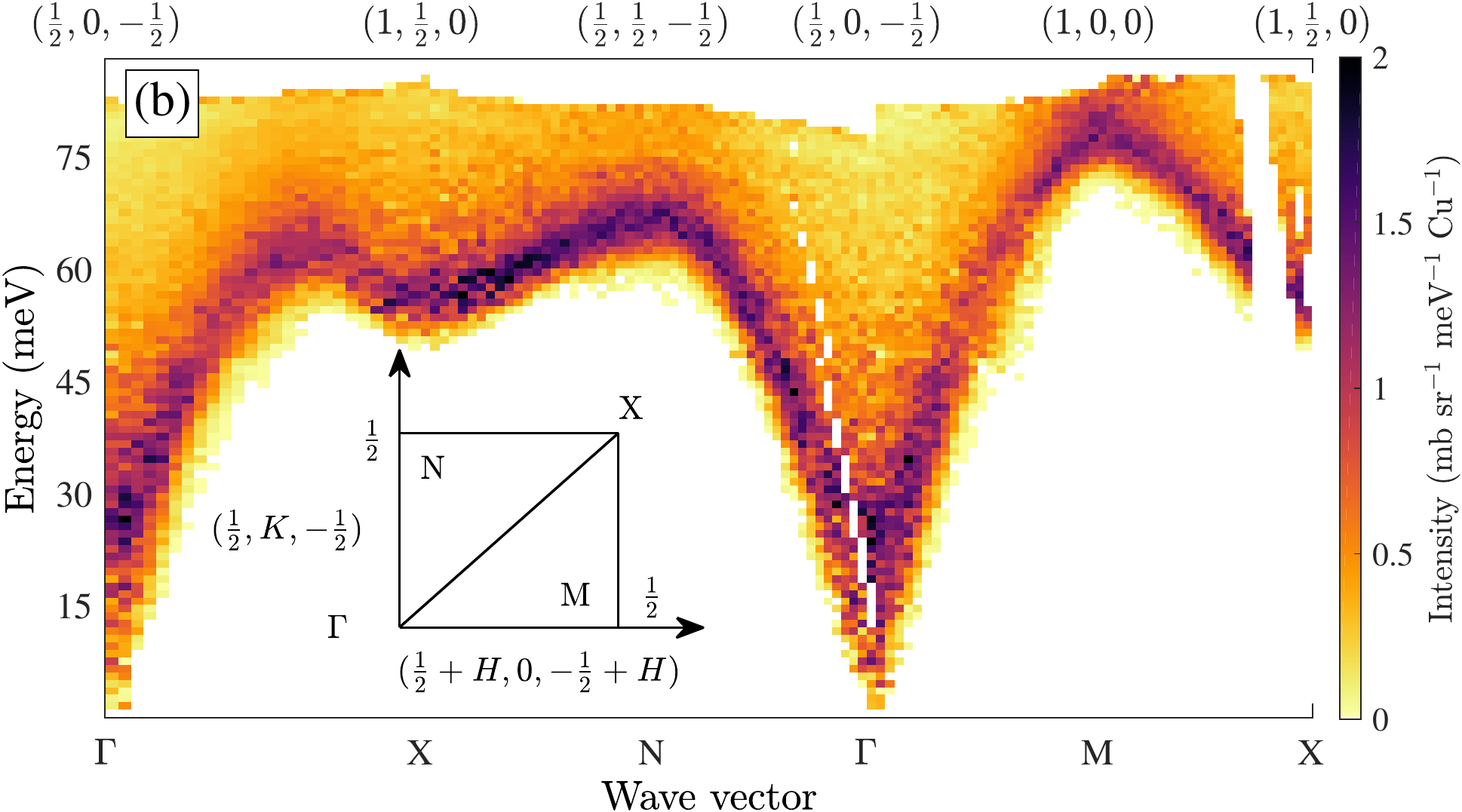}
 \caption{The magnetic dispersion of CuO in the AF1 phase at 6~K. (a) Neutron scattering data from the MAPS spectrometer, showing intensity in units of mb sr$^{-1}$ meV$^{-1}$ Cu$^{-1}$ as a function of energy transfer and wave vector. (b) Our linear spin-wave model convolved with the resolution function of MAPS. The measured background scattering is not constant and has not been included in the simulation. The  intensity scale in (b) is therefore different from that in (a). The observed and calculated spectra are in good agreement, with a goodness-of-fit parameter $\chi^2 = 2.6$. The inset shows the path in reciprocal space along which the dispersion is plotted. Here, $\Gamma$ represents the AFM wave vector $\textbf{q}_1 = (\frac{1}{2},0,-\frac{1}{2})$. 
}
 \label{fig:AF1_neutron_map}
\end{figure}

Next, we investigate the magnetic scattering in the inter-chain directions. Fig.~\ref{fig:AF1_neutron_map}(a) shows the intensity of scattered neutrons as a function of energy and wave vector along the path $\Gamma \rightarrow \textrm{X} \rightarrow \textrm{N} \rightarrow \Gamma \rightarrow \textrm{M} \rightarrow \textrm{X}$ in the Brillouin zone, given in the inset of ~Fig.~\ref{fig:AF1_neutron_map}(b).  The low ($E<48$~meV) and high energy ($E>48$~meV) parts of the spectrum were obtained in different Brillouin zones, and in some regions data from different Brillouin zones were combined  to increase statistics.
In previous work only  the dispersion along the $\Gamma \rightarrow \textrm{M}$ and $\Gamma\rightarrow \textrm{N}$ directions has been investigated \cite{Ain1989}.

This inter-chain spectrum shows a well defined spin-wave dispersion with a band width of around 80~meV, which indicates non-negligible inter-chain interactions. At the X point there is an intriguing softening of the dispersion, and at the $\Gamma$ point there is a broadening of the spectrum as a function of energy which we have studied in more detail by polarised neutron inelastic scattering. Strong phonon scattering is seen around 15\,meV throughout the Brillouin zone.

\begin{figure}
 \centering
 \includegraphics[width=0.22\textwidth]{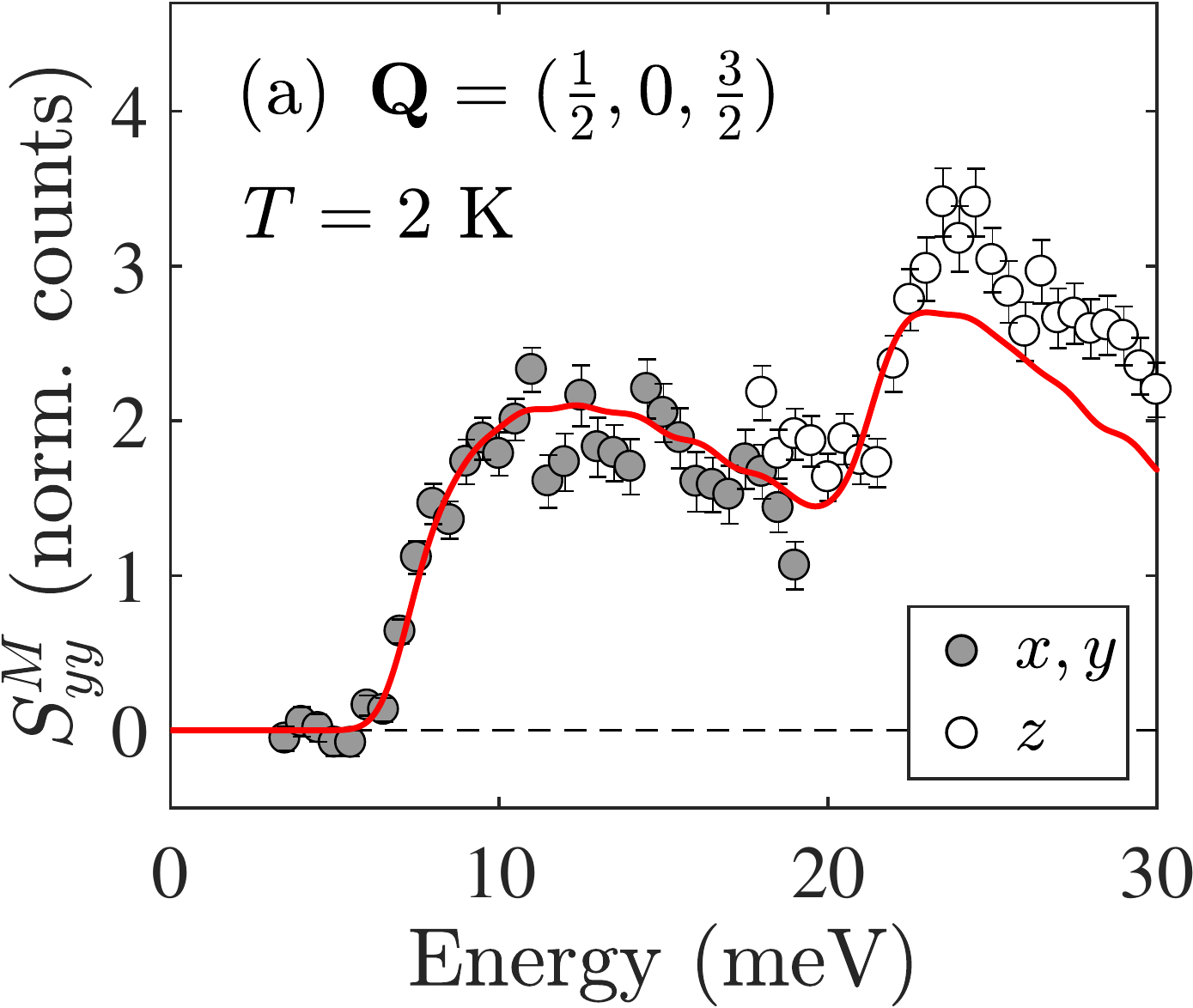}
 \includegraphics[width=0.22\textwidth]{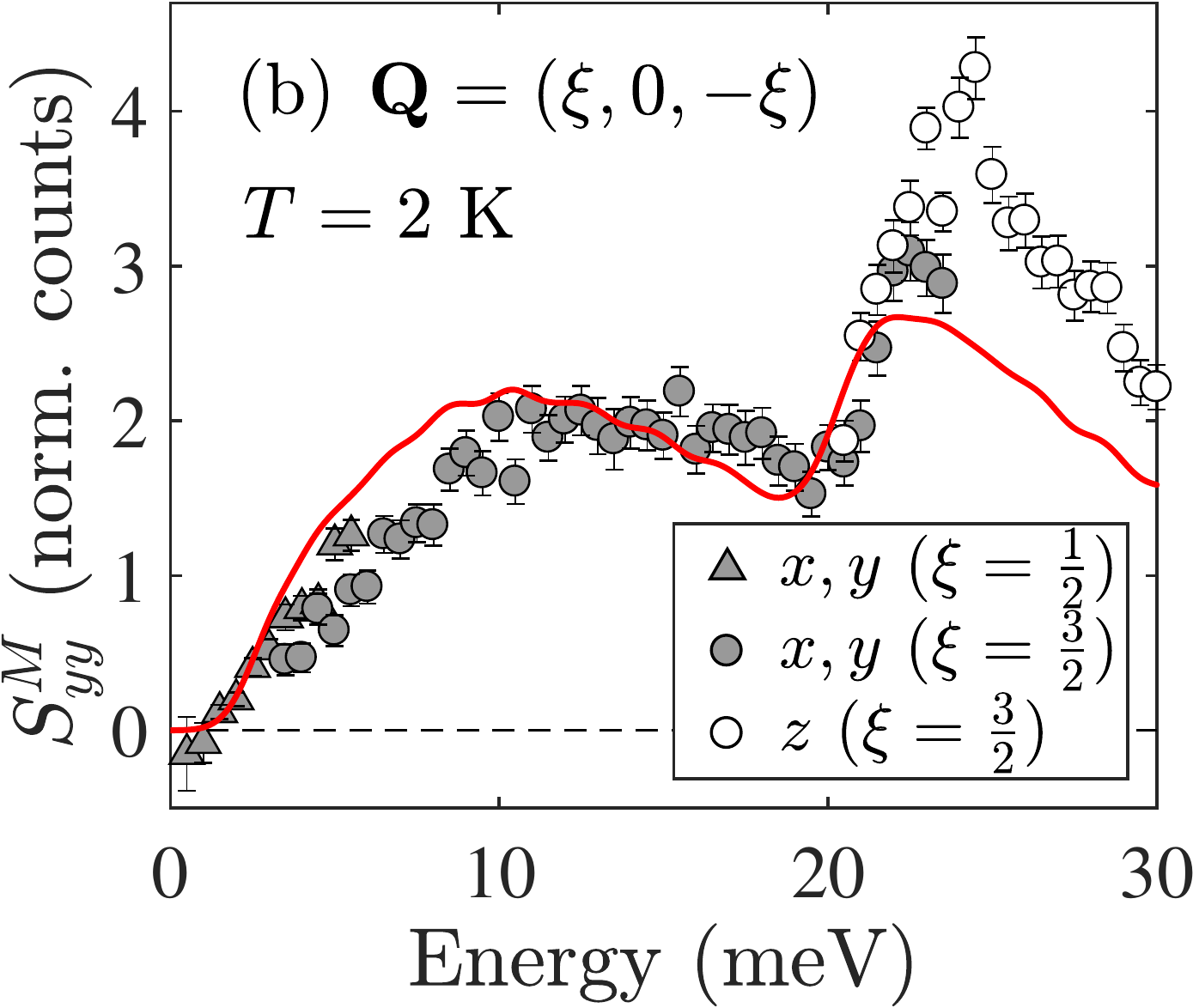}
 \caption{Magnetic response function $S^M_{yy}$ of CuO measured at AF1-phase AFM zone centres by neutron polarisation analysis. (a) $\textbf{Q} = (\frac{1}{2},0,\frac{3}{2})$. (b)  $\textbf{Q} = (\frac{1}{2},0,-\frac{1}{2})$ and $(\frac{3}{2},0,-\frac{3}{2})$. The filled symbols show   $S^M_{yy}$ deduced from the $x$ and $y$ spin-flip (SF) scattering, Eq.~(\ref{eq:Ix-Iy}), and the open symbols are from the SF scattering in the  $z$  channel after subtraction of an estimate of the SF  background and nuclear spin-incoherent scattering, Eq.~(\ref{eq:IzSF}). Data have been corrected for the experimental elastic magnetic structure factors and Cu magnetic form factor reported in Ref.~\onlinecite{Forsyth1988}. The red line is our LSWT model convolved with an approximation to the resolution function of the instrument. 
 } 
\label{fig:neutron_Iyy_lowT}
\end{figure}

In Fig.~\ref{fig:neutron_Iyy_lowT} we show magnetic scattering as a function of energy at AFM zone centres located along two distinct directions in $\textbf{Q}$: $\QQQ_1=(\frac{1}{2},0,\frac{3}{2})$ and $\QQQ_2=(\frac{1}{2},0,-\frac{1}{2})$. The angle between $\QQQ_1$ and $\QQQ_2$ is $108^\circ$.   The magnetic signal has been obtained by neutron polarisation analysis on the IN20 triple-axis spectrometer and corresponds to the magnetic response function $S^M_{yy}$, given by
\begin{equation}
S^M_{yy}(\QQQ,\omega) = \frac{1}{2\pi\hbar} \int_{-\infty}^\infty  \langle M_y^\dagger(\QQQ)  M_y(\QQQ,t)\, \rangle\, \textrm{e}^{-\textrm{i} \omega t}\,\mathrm{d}t.
\end{equation}
Here, the Blume--Maleev coordinate system is used,  in which $x \parallel \QQQ$, $y \perp \QQQ$ in the horizontal $a$--$c$ plane, and $z$ is vertical ($\parallel b$). Therefore, $S^M_{yy}$ measures spin fluctuations that are perpendicular to both  $\textbf{Q}$ and the $b$ axis.

At both positions there is a low energy gap and an asymmetric peak at an energy of about 23~meV. The gap is anisotropic, changing from about 7~meV at  $(\frac{1}{2},0,\frac{3}{2})$ [Fig.~\ref{fig:neutron_Iyy_lowT}(a)]  to less than 2~meV at $(\frac{1}{2},0,-\frac{1}{2})$ [Fig.~\ref{fig:neutron_Iyy_lowT}(b)].  The step-like features in these energy scans are the result of the spectrometer resolution scanning through the minima in the highly dispersive magnon bands.

We now present a minimal model that explains all the key features of the data highlighted above. We model the spin dynamics using linear spin-wave theory (LSWT), which has been shown to give surprisingly accurate results for 3D ordered magnets, even those with $S=\frac{1}{2}$ (Ref.~\onlinecite{Anderson1952}).

To model the magnetic anisotropy we make the interaction along the $[10\bar{1}]$ chains, $J$, anisotropic, setting $J_{xx} = J$, $J_{yy} = J-D_1$, $J_{zz}= J-D_2$ and all off-diagonal terms zero.
The coordinate system here has $x$ along $\vvv$, $y$ along $\bbb$ and $z$ completes the right-handed set. It is seen that $z$ is perpendicular to the plane of rotation of the spins in the AF2 phase. $D_1<0$ makes the $b$ axis an easy axis, while $D_2>0$ makes the $b$--$v$ plane an easy plane for the spins. \footnote{In the mean field approximation this approach is nearly identical to using effective single-ion anisotropies: The $J$ term of the Hamiltonian can be written $\sum_\text{chains} \sum_i J S_i^x S_{i+1}^x + (J-D_1) S_i^y S_{i+1}^y+ (J-D_2) S_i^z S_{i+1}^z = \sum_\text{chains} \sum_i [J \SSS_i \cdot \SSS_{i+1} + D_1 (S_i^y)^2 + D_2 (S_i^z)^2]$, where we have used that along the chains in the AF1 phase (and approximately in the AF2 phase) $\SSS_i = -\SSS_{i+1}$.}

When $D_2=0$ the dispersion is straightforward to calculate analytically in LSWT and can reproduce all the features of the observed spectrum apart from the anisotropic gap at the antiferromagnetic zone centre. We give the formula in Appendix~\ref{sec:analytical}. For the more general case of $D_2\neq 0$ we use the program SpinW \cite{Toth2015} to diagonalize the Hamiltonian numerically.

The magnetic unit cell of CuO in the AF1 phase contains eight spins [Fig.~\ref{fig:exchange_interactions}(a)], so there are eight spin-wave modes for a given wave vector. However, with the translational symmetry introduced by setting $J_{ab}=J_{ab}'$ the modes become doubly degenerate, leaving four distinct pairs of modes\footnote{When $J_{ab}=J_{ab}'$ a unit cell of half the size can be used to describe the magnetic order and excitations.}. As mentioned above, the spins divide into two antiferromagnetic lattices which in the ideal structure with isotropic Heisenberg interactions are fully frustrated with respect to one another. If the frustration is relieved then the four pairs of modes split into two quadruplets, corresponding to acoustic and optic spin waves, respectively.  The fourfold degeneracy of the optic and acoustic modes is lifted by easy-plane anisotropy ($D_2 > 0$) giving two non-degenerate pairs of acoustic modes and two non-degenerate pairs of optic modes, and the lowest energy (Goldstone) pair of modes is gapped at the AFM zone centre due to the axial anisotropy ($D_1 < 0$).

The sharp increase in the $S_{yy}^M$ magnetic response at 23~meV (Fig.~\ref{fig:neutron_Iyy_lowT}) is almost identical for the two approximately perpendicular wave vectors and can be identified with the onset of the group of four nearly degenerate optic modes. The difference between the magnetic response at the two wave vectors observed for energies below the optic modes implies a splitting of the acoustic modes by easy-plane anisotropy. This interpretation is confirmed by the 7~meV gap in the scan in Fig.~\ref{fig:neutron_Iyy_lowT}(a) at $\textbf{Q} = (\frac{1}{2},0,\frac{3}{2})$. This $\textbf{Q}$ is parallel to the easy plane, so the magnetic scattering here arises purely from spin fluctuations perpendicular to the plane, which are absent below 7~meV. For $\textbf{Q}$ positions along $(\xi,0,-\xi)$ [Fig.~\ref{fig:neutron_Iyy_lowT}(b)] the $S_{yy}^M$ channel mainly probes spin fluctuations in the easy plane. These are seen to extend well below 7~meV. The small gap of order 1~meV seen in Fig.~\ref{fig:neutron_Iyy_lowT}(b) indicates a small easy-axis anisotropy within the easy plane.

\begin{figure}[t]
\centering
 \includegraphics[height=0.13\textheight]{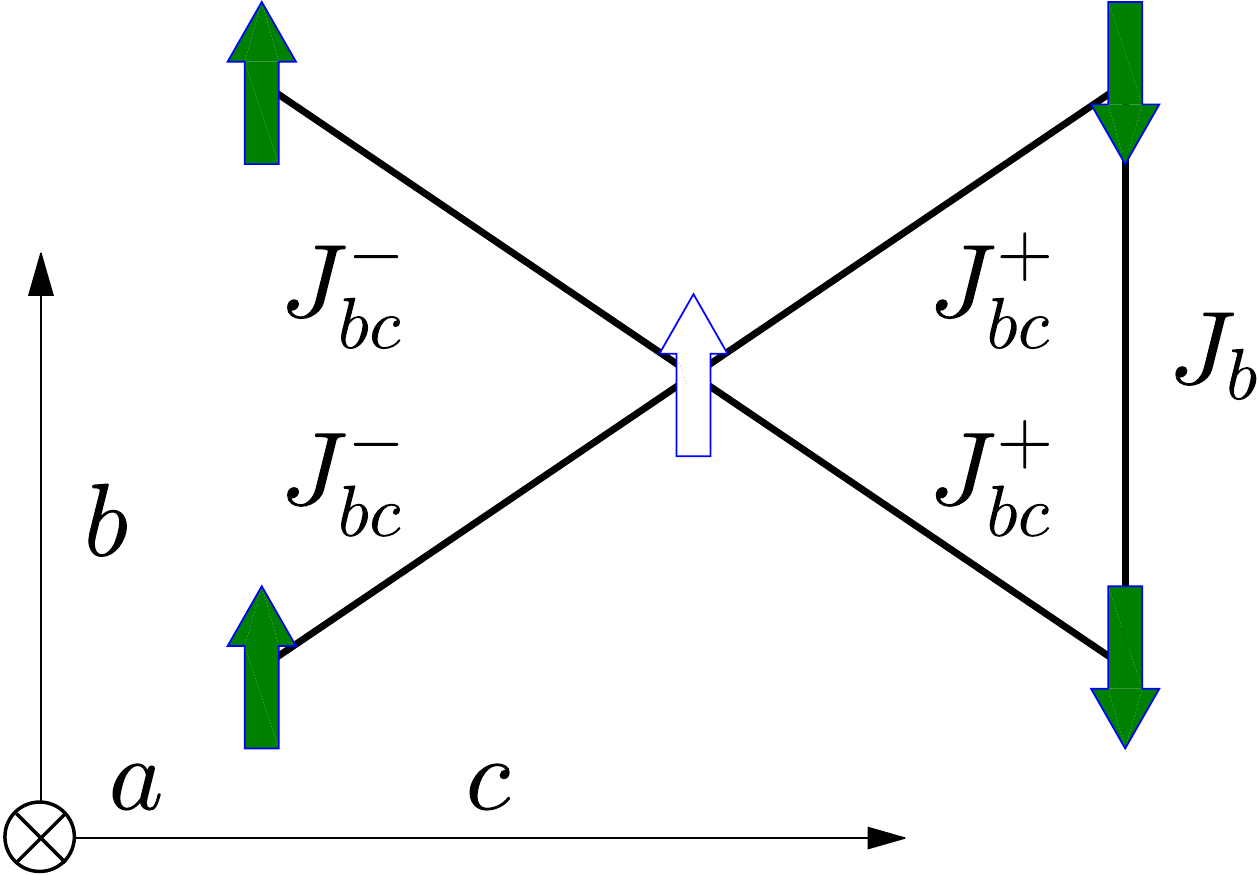}
 \caption{ Projection down the $b$ axis onto the $bc$ plane, showing the assumed $J_{bc}^- \ne J_{bc}^+$ introduced in our model to relieve the frustration.
 }
 \label{fig:J_bc_split}
\end{figure}

The exchange interactions included in our model have been described in Section~\ref{sec:structure} and are defined in Fig.~\ref{fig:exchange_interactions} and Table~\ref{tab:exchange_paths}. The observed splitting of the optic and acoustic modes at the AFM zone centre implies that an unbalanced coupling exists between the two AFM lattices, whereas in the ideal AF1 structure with Heisenberg interactions the coupling is fully frustrated (see Fig.~\ref{fig:exchange_interactions}).  Such an imbalance could arise from a subtle structural distortion to a lower symmetry. We shall return to this point later, but for the time being we shall relieve the frustration between the two lattices by splitting the  $J_{bc}$ couplings such that the Cu sites are no longer at a centre of symmetry of the magnetic structure, see Fig.~\ref{fig:J_bc_split}.  We define
\begin{equation}
\begin{aligned}
J_{bc}^- & = J_{bc}-\delta/2, \\[2pt]
J_{bc}^+ & = J_{bc}+\delta/2,
\end{aligned}
\label{eq:delta}
\end{equation}
so that $\delta = 0$ is fully frustrated and relief of frustration increases with $|\delta|$. We note that this lifting of the frustration could not have been achieved with $J_{ab}'$ or $J_{bc}'$ mentioned above.

The dominant exchange interaction $J$ has already been determined from the high energy part of the spectrum (see above) and it was treated as a fixed parameter in subsequent fits to the inter-chain dispersion. As LSWT treats spin semi-classically it cannot capture the full spinon dynamics of the $S=\frac{1}{2}$ Heisenberg AFM chain and the value of $J$ must be renormalised \cite{Coldea2002}. We do this by defining an effective exchange parameter $J^\text{sw}$ which produces a spin-wave dispersion in LSWT that matches the lower bound of the spinon continuum Eq.~(\ref{eq:1d_dispa}) for an exchange parameter $J$. The correspondence between these parameters is  $J^\text{sw}=\pi J/2$. Hence, the value $J=91.4$~meV obtained in Section~\ref{sec:results} from a fit to the two-spinon dispersion translates to $J^\text{sw} = 143.6$~meV, and so this is the value we used in our LSWT fits to the inter-chain spin-wave spectrum.

We fitted the spin-wave model by a least-squares algorithm to the observed spin-wave spectrum along several symmetry directions in the Brillouin zone, as shown in Fig.~\ref{fig:AF1_neutron_map}(b). The procedure was to take a number of constant-$\QQQ$ cuts through the raw data and fit these simultaneously to the resolution-convolved spin-wave spectrum.  SpinW was used to calculate the spectrum, and the resolution convolution and fitting were performed by the Horace and Tobyfit software\cite{Ewings2016}. In addition to the exchange constants, a global amplitude factor and a separate background parameter for each cut were varied. The background is not included in Fig.~\ref{fig:AF1_neutron_map}(b), and so the intensity scale is not identical to the experimental data in Fig.~\ref{fig:AF1_neutron_map}(a).

Ignoring the data at the $\Gamma$ point for the moment, we obtained a good fit using only the exchange constants $J$, $J_{ac}$, $J_b$, and $J_{2,ab}$. The quality of the fit did not improve when the second-neighbour parameters $J_{2,a}$ and $J_{2,c}$ were allowed to vary, and from now on these will be set to zero.  We remark that inclusion of the spectrometer resolution is quite important for a quantitative model. Fits performed with and without resolution convolution returned parameters which differed by up to 20\%. 

The structure in the intensity observed at the $\Gamma$ point (Fig.~\ref{fig:neutron_Iyy_lowT}) is largely determined by $\delta$, $D_1$ and $D_2$. To find the best fit to the data, these parameters were varied while keeping $J$, $J_{ac}$, $J_b$ and $J_{2,ab}$ fixed. 
The intensity was calculated by convolving the model with a four-dimensional Gaussian function $R(E,\textbf{Q})$ with full width at half maxima (FWHM) of $\Delta E = 1.5$~meV along the energy axis and $\Delta \textbf{Q} = (0.13, 0.17, 0.13)$~r.l.u. along the $a^*$, $b^*$ and $c^*$ axes to approximate the resolution of the IN20 spectrometer. This procedure models quite well the resolution broadening of the line shape but does not take into account changes in the orientation and volume of the resolution function with energy. It is adequate, therefore, for extracting the energies of the mode onsets but is not expected to describe the scattering intensity accurately. To improve the intensity calculation would require a description of the instrument which goes far beyond standard resolution models, and which is not currently available.

The fit is shown as the red line in Fig.~\ref{fig:neutron_Iyy_lowT}, and is seen to be in good agreement with the data, especially for energies below 20~meV. Most importantly, we find excellent agreement with the  onset energies of the three observed modes ($\sim$1~meV, 7~meV and 23~meV).  These energies determine the three parameters,  $\delta$, $D_1$ and $D_2$, and are essentially independent of the model for the resolution function.  The discrepancy in intensity between the data and model at higher energies is likely partly because a different experimental setup was used to record the higher energy data (different $k_\textrm{f}$ and polarization channel) and partly because of the limitations of our model for the resolution function, as explained above.

The best fit to our data was obtained with the model parameters listed in Table~\ref{tab:model_parameters} which are seen to provide a good description of the complete measured spin-wave spectrum.

We note that $J_{ab}$ and $J_{bc}$ have little effect on the dispersion, and thus cannot be determined by modeling the inelastic neutron scattering data presented in this section. However, $J_{ab}$ and $J_{bc}$, together with $\delta$, are important for selecting the ground state, as discussed in the next section.

\subsection{Spin dynamics in the AF2 phase}

\begin{figure}
 \centering
  \includegraphics[width=0.22\textwidth]{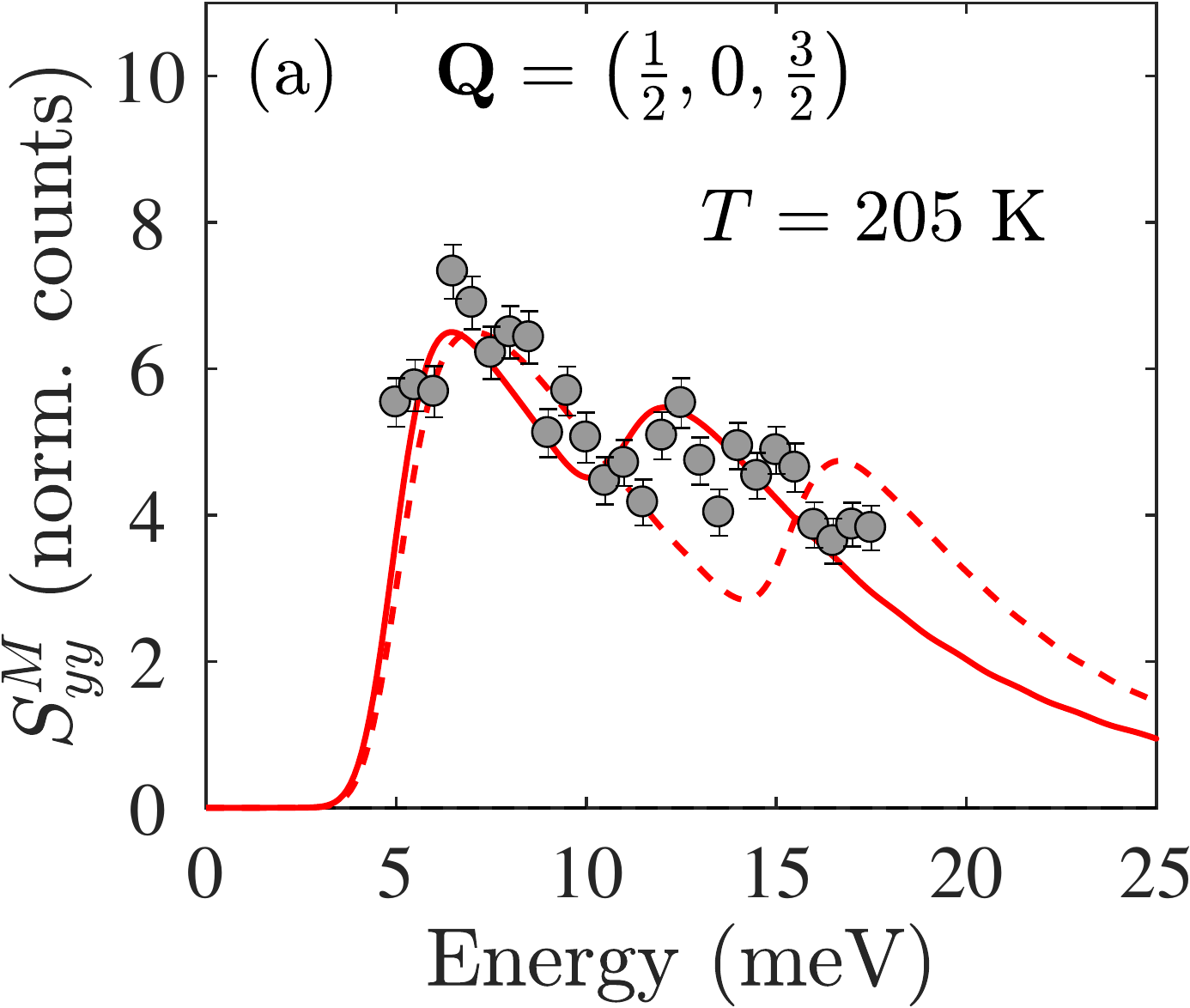}
  \includegraphics[width=0.22\textwidth]{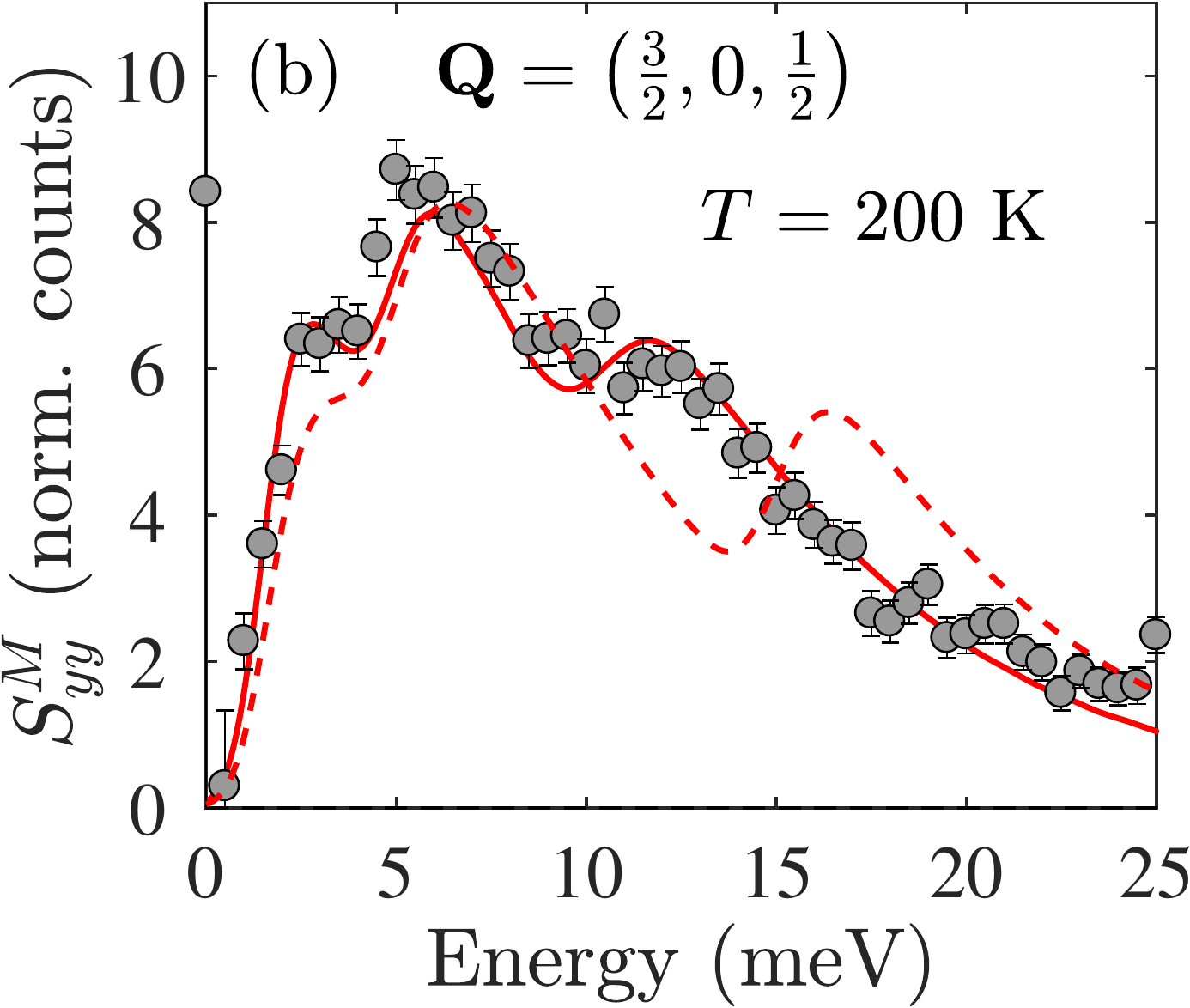} \\
  \includegraphics[width=0.22\textwidth]{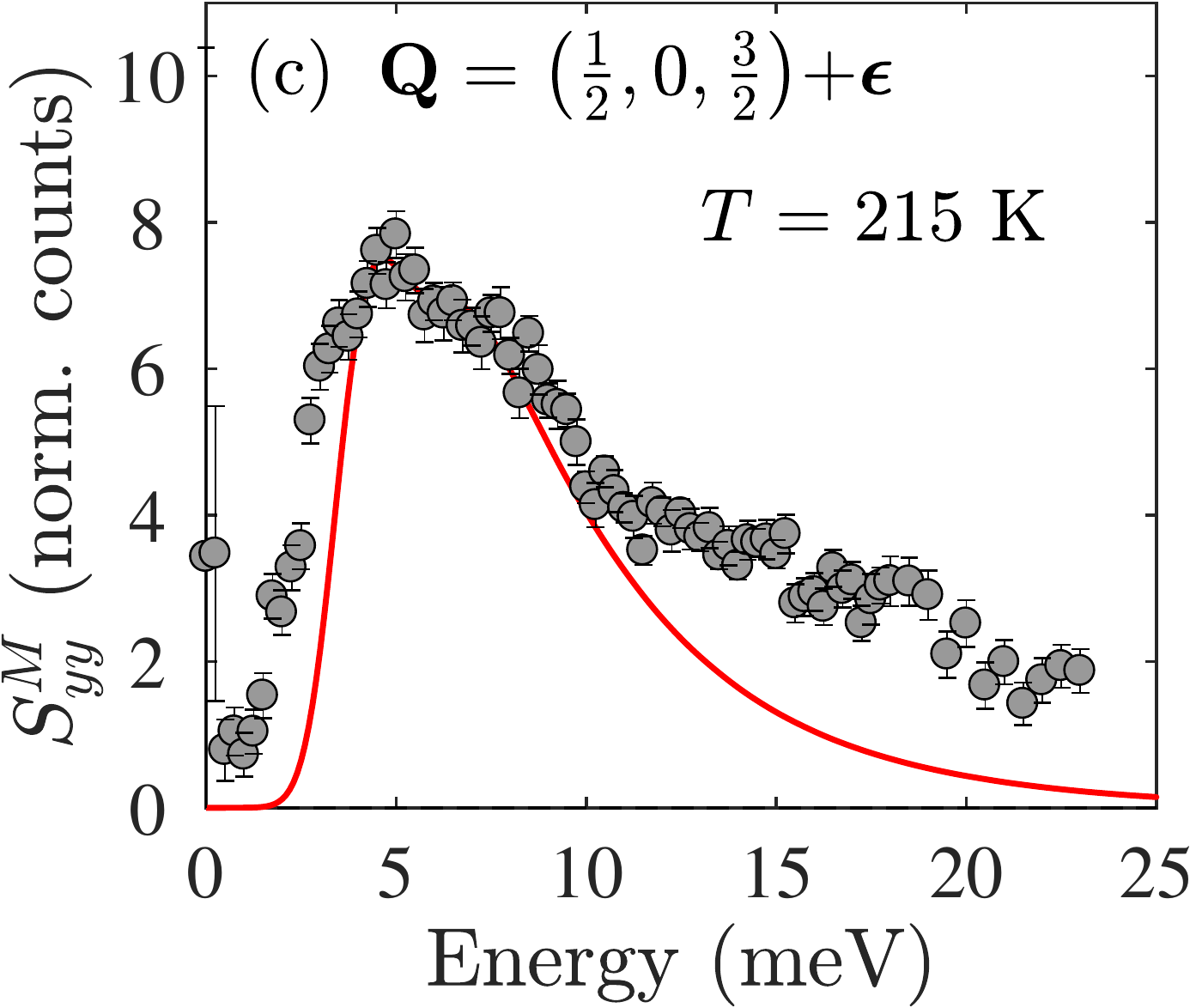}
  \includegraphics[width=0.22\textwidth]{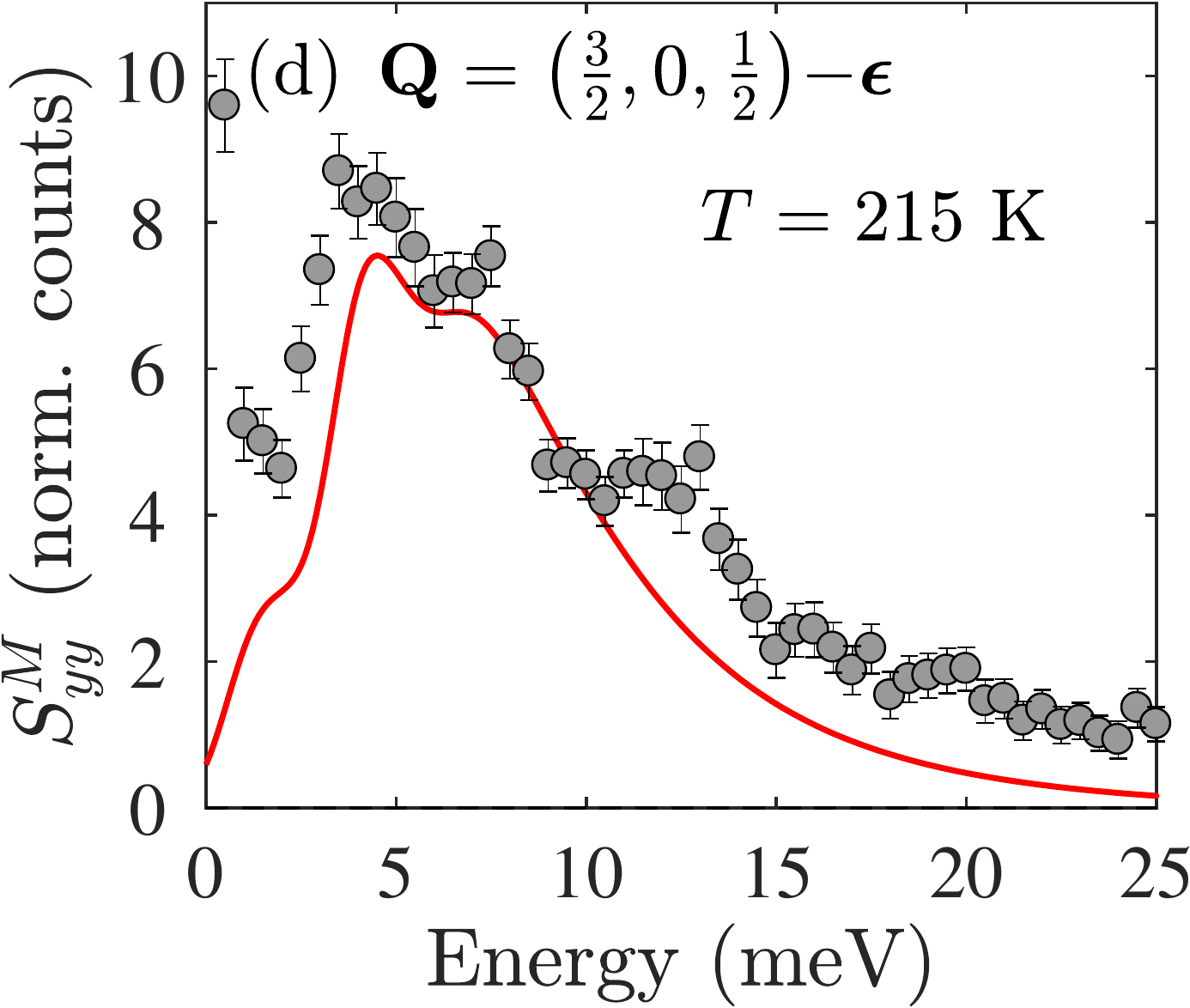} \\
  \includegraphics[width=0.22\textwidth]{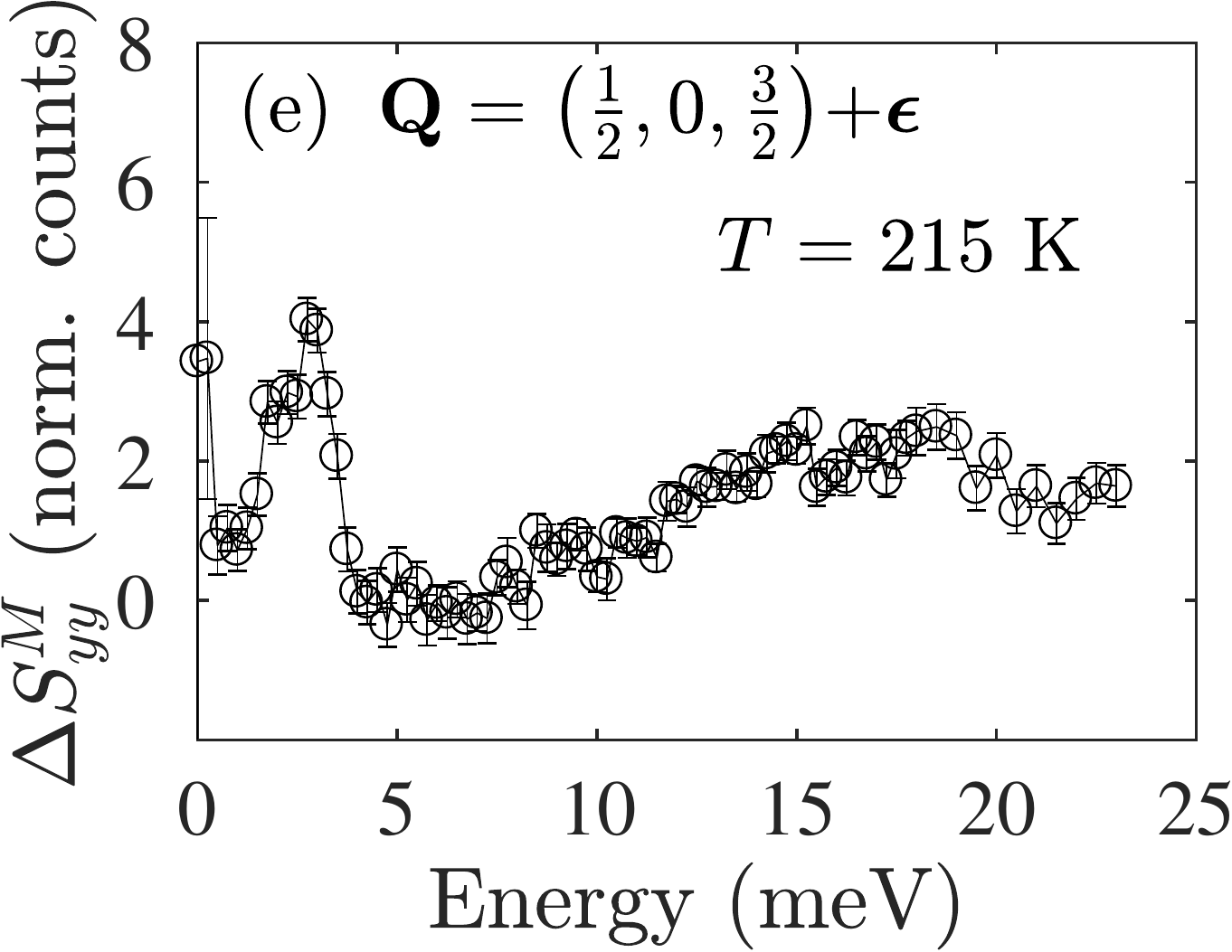}
  \includegraphics[width=0.22\textwidth]{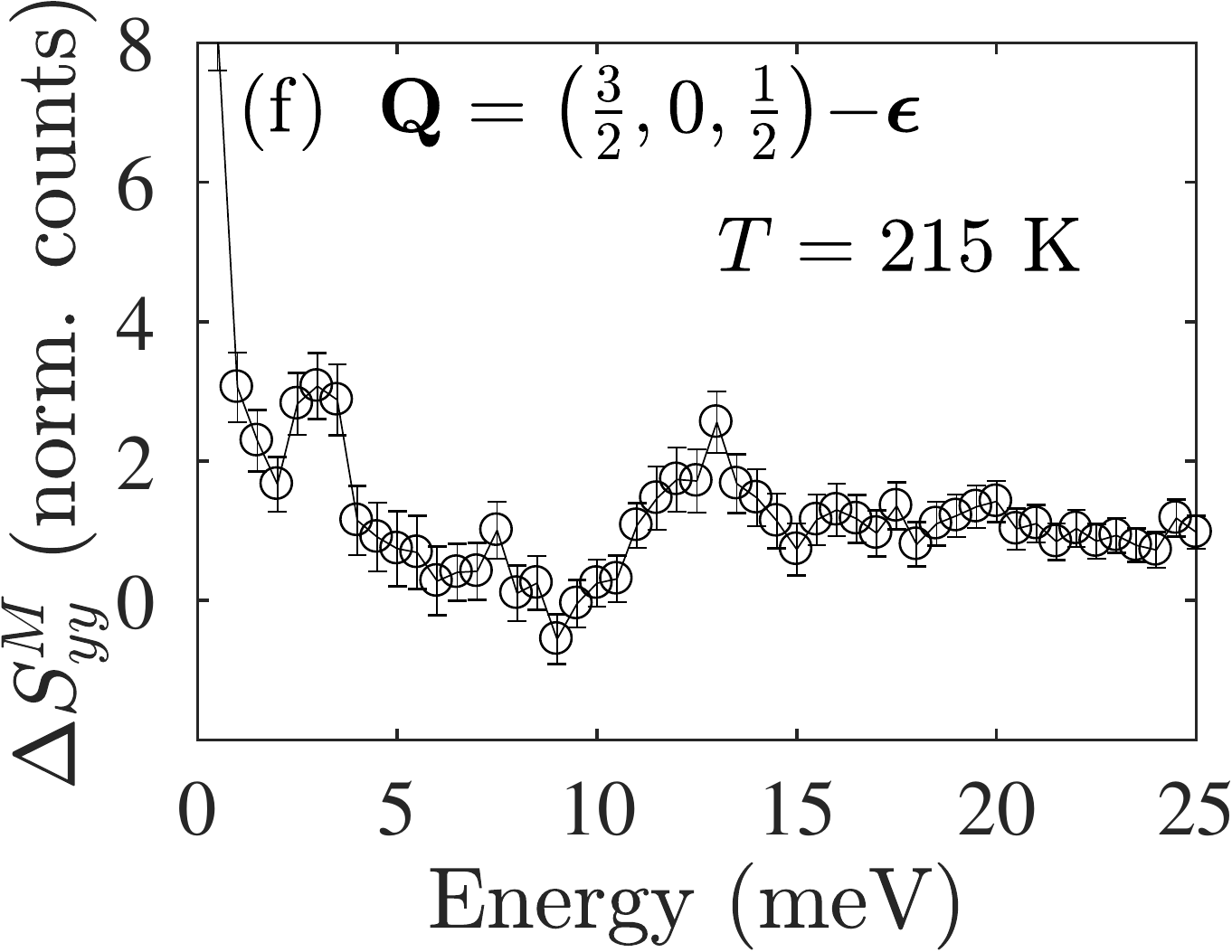}
 \caption{
 Magnetic response function $S^M_{yy}$ of CuO measured at elevated temperatures by neutron polarisation analysis.  (a)--(b) AF1-phase AFM zone centres at $T \simeq 200$~K. (c)--(d) AF2-phase magnetic positions at $T = 215$~K. (e)--(f) Difference between the data and model in (c) and (d).  The solid red lines in (a)--(d) are calculated from our LSWT model convolved with an approximation of the resolution function of the instrument. The dotted lines in (a) and (b) show the model with $\delta=0.68$~meV as found for the low temperature data.
}
 \label{fig:compare_AF2_yy}
\end{figure}

We now investigate the magnetic dynamics at elevated temperatures. To begin with, we look at the AF1 phase at high temperature. Figures~\ref{fig:compare_AF2_yy}(a) and (b) show the $S_{yy}^{M}$ magnetic scattering at $\sim$200~K as a function of energy at two AF1-phase AFM zone centres located along distinct directions in $\textbf{Q}$. At $\QQQ=(\frac{3}{2},0,\frac{1}{2})$, Fig.~\ref{fig:compare_AF2_yy}(b), we observe three clear features: a peak centred near 6~meV and shoulders at about 3 and 12~meV. At $\QQQ=(\frac{1}{2},0,\frac{3}{2})$, Fig.~\ref{fig:compare_AF2_yy}(a), we also observe a peak at 6~meV and a shoulder at about 12~meV (the data do not extend to energies below 5~meV).

Compared with the low temperature data shown in Fig.~\ref{fig:neutron_Iyy_lowT}, the onset energies of the optic and acoustic modes at $\sim$200~K have been significantly reduced. If we only take into account the observed decrease in the AFM-ordered moment with increasing temperature\cite{Forsyth1988}, and assume the low temperature model parameters, then at 200~K the acoustic mode onsets are expected to be at about 1.4~meV and 4.8~meV (split due to the anisotropy), and the optic mode onset at about 15~meV. The former agree well with the high temperature data (the lower mode is more prominent at elevated temperature due to thermal population)  while the latter is close to, but higher than, the feature at 12~meV. As there is no feature in the $\sim$200~K data at 15~meV we identify the 12~meV signal with the optic mode, from which it follows that the inter-lattice net coupling must be smaller at 200~K than at 2~K.  We find that the $\sim$200~K data is described quite well by the low-temperature model providing $\delta$ is reduced from 0.68~meV to 0.3~meV. Simulations of resolution-convolved model spectra are shown in Fig.~\ref{fig:compare_AF2_yy}(a) and (b).

We now turn to the AF2 phase. Figures~\ref{fig:compare_AF2_yy}(c) and (d) show  measurements of $S_{yy}^M$ at 215~K. At $\QQQ=(\frac{1}{2},0,\frac{3}{2})+\mbox{\boldmath $ \epsilon$}$ we see broad features at 3, 5 and 8~meV and no elastic scattering, while at $\QQQ=(\frac{3}{2},0,\frac{1}{2})-\mbox{\boldmath $ \epsilon$}$ we see broad features at 3, 5, 8 and 13~meV, and a strong elastic signal.

The Blume--Maleev $y$ direction for $(\frac{1}{2},0,\frac{3}{2})+\mbox{\boldmath $ \epsilon$}$ is nearly perpendicular to the plane of the helix, and the absence of an elastic signal in Fig.~\ref{fig:compare_AF2_yy}(c) shows that, as  expected, there is no ordered magnetic moment in this direction. Conversely, the $y$ direction for $(\frac{3}{2},0,\frac{1}{2})-\mbox{\boldmath $ \epsilon$}$ has a significant component in the plane of the helix, which is why an elastic signal is observed here.

\begin{figure}
\centering
\includegraphics[width=0.22\textwidth]{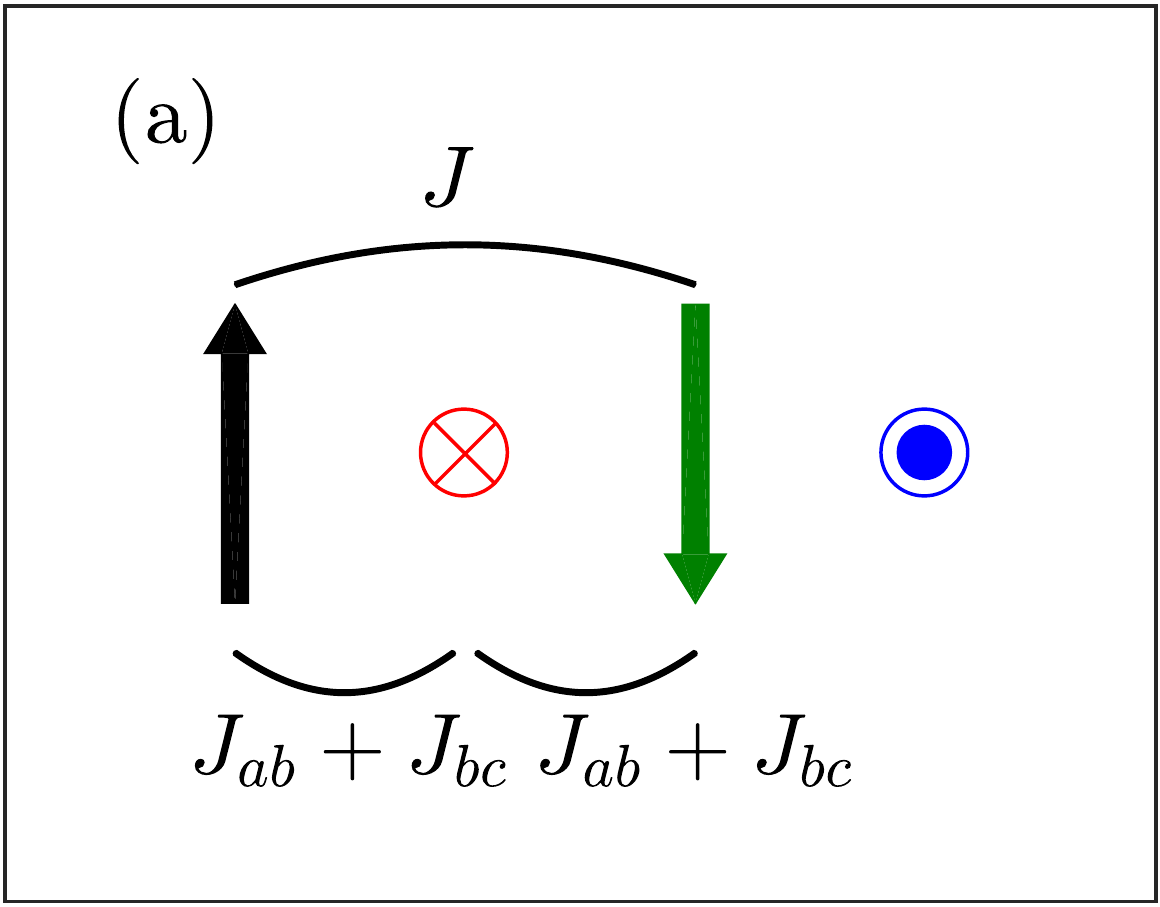}
\includegraphics[width=0.22\textwidth]{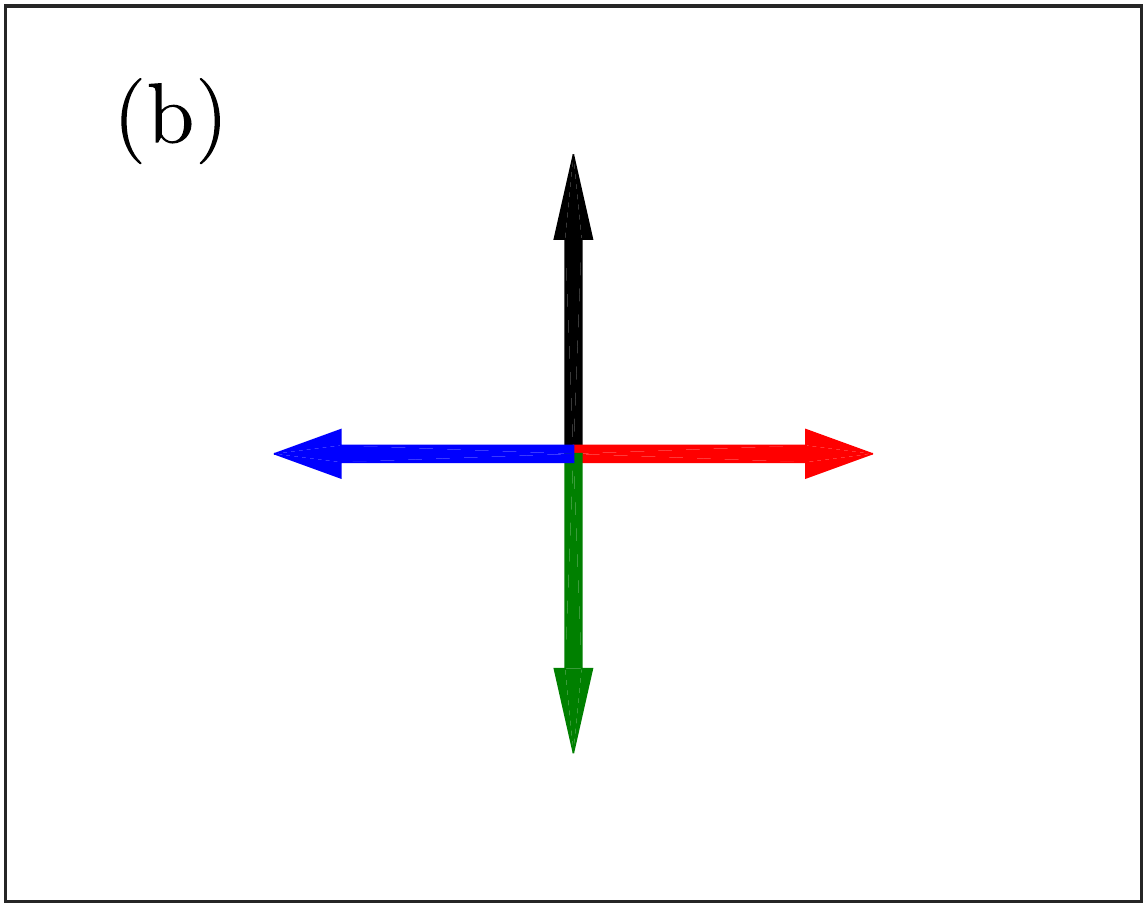}\\
\includegraphics[width=0.22\textwidth]{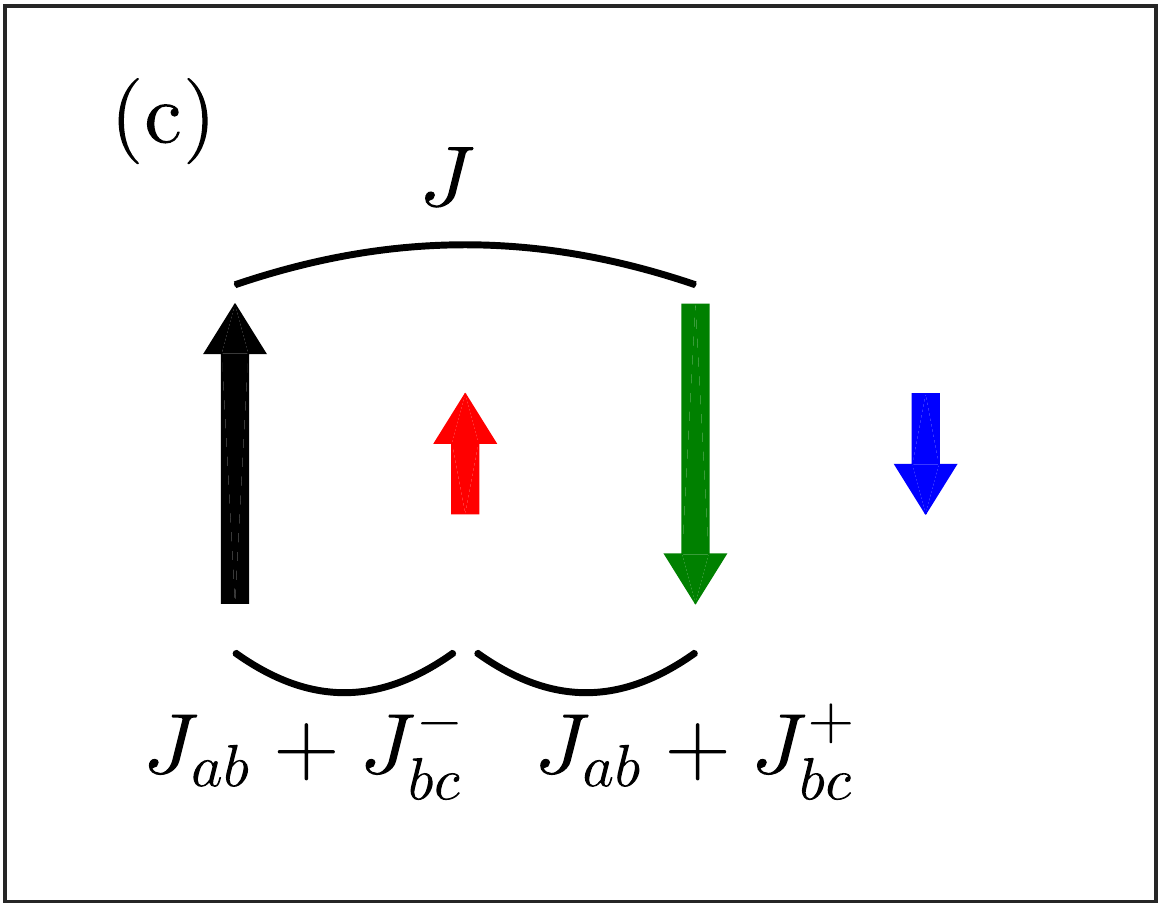}
\includegraphics[width=0.22\textwidth]{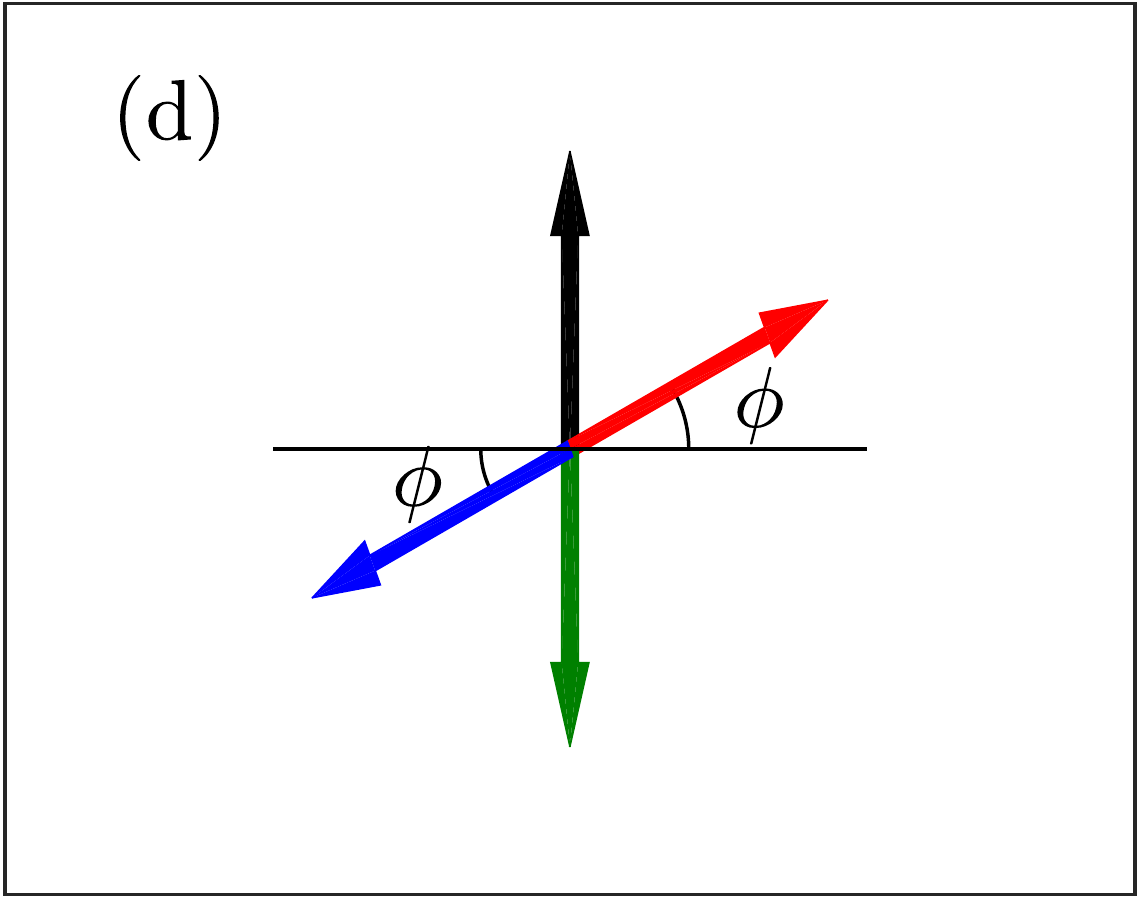}
\caption{Schematic of the AF2 phase helicoidal spin structure projected onto a single chain. The chain can be described by the $J_1$--$J_2$ model, where $J_1$ is the coupling between nearest-neighbor spins, and $J_2 (=J)$ is the next-nearest neighbor coupling. (a) and (b) shows the spin structure for $\delta=0$ as viewed from the side and along the chain, respectively. (c) and (d) show how two of the spins rotate by an angle $\phi$ when $\delta$ is non-zero. For the sake of simplicity we have taken the plane of rotation of the spins to be perpendicular to the chain.}
\label{fig:helix_illustration}
\end{figure}

In order to model the AF2 phase we must find a set of exchange and anisotropy parameters which stabilizes the helicoidal spin structure, then calculate the magnetic spectrum and compare it with the experimental data at low energies shown in Figs.~\ref{fig:compare_AF2_yy}(c)--(f).  To understand the stability of the AF2 spin arrangement in CuO it might be helpful to view it as a collection of $J_1$--$J_2$ spin chains running along the $[1 0 \overline{1}]$ direction, where $J_1$ and $J_2$ are nearest and next-nearest neighbour couplings --- see Fig.~\ref{fig:helix_illustration} and Fig.~\ref{fig:exchange_interactions}(d).  The $J_1$--$J_2$ model is known to produce a helical structure with a propagation vector $q_\textrm{ch}$ given in mean-field theory by $\cos(2 \pi q_\textrm{ch}) = -J_1/4J_2$. In our model $J_2 \equiv J$ and, for $\delta = 0$, $J_1 \equiv J_{ab}+J_{bc}$. For CuO, therefore, $|J_1| \ll |J_2|$, and for $\delta = 0$ we expect a helix with a propagation vector $q_\textrm{ch}\simeq 0.5$, i.e.~an angle of approximately 90$^\circ$ between neighbouring spins, as reported.

When $\delta \ne 0$ there are two different $J_1$ values, given by $J_{ab}+J_{bc}^-$ and $J_{ab}+J_{bc}^+$. Neighbouring spins coupled by $J_{bc}^-$ rotate towards one other so the angle between them is $\simeq 90^\circ -\phi$, while the angle between neighbouring spins coupled by $J_{bc}^+$ is $\simeq 90^\circ +\phi$, see Figs.~\ref{fig:helix_illustration}(c) and (d). At some critical value of $\delta$ the AF1 collinear magnetic structure becomes the stable phase ($\phi = 90^\circ$).

In our analysis of the AF2 phase we fixed the exchange parameters $J$, $J_{ac}$, $J_{2,ab}$ and $J_b$ to their values determined from the low temperature measurements, Table~\ref{tab:model_parameters}, and we fixed the AF2 propagation vector $\textbf{q}_2$ to the observed wave vector $\qqq_2 = (0.506,0,-0.483)$. The planar anisotropy has no impact on the calculations, except that it selects the plane in which the spins rotate. We neglect the small axial anisotropy parametrized by $D_1$. 
We investigated both analytic and numerical solutions of the mean-field equations for the spin structures in CuO --- details are given in Appendix~\ref{sec:helix_calculation}. For fixed $\qqq_2$, the parameters $J$, $J_{ac}$, $J_{2,ab}$ and $\delta$ uniquely determine the values of $J_{bc}$, $J_{ab}$ and $\phi$. The value of $\phi$ depends strongly on $\delta$.

Despite an extensive search, including $J_{2,a}$, $J_{2,c}$ and the Dzyaloshinskii--Moriya interaction (see below), we were not able to achieve a complete description of the low energy neutron scattering data in the AF2 phase. The curves drawn on Figs.~\ref{fig:compare_AF2_yy}(c) and (d) show the best match that we could find. To achieve this we needed to assume that the optic mode shifts down from 12~meV at $\sim$200~K to 8~meV at 215~K. This shift is reproduced via a reduction in the $\delta$ parameter to $\delta=0.15$~meV, which gives $\phi=32^\circ$. The simulated lineshapes then agree quite well with the data below $\sim$10~meV, but there are deviations at higher energies. Some of these deviations will be due to inadequacies in the simple model we used for the spectrometer resolution, but it is notable that the simulation is not able to reproduce the feature observed at 12~meV in Fig.~\ref{fig:compare_AF2_yy}(d). It is possible that this feature is the optic mode, as observed at $\sim$200~K, but in that case it should also be present in the scan in Fig.~\ref{fig:compare_AF2_yy}(d), which does not appear to be the case. Simulations in which the optic mode was fixed at 13~meV gave poor results for the line shape below $\sim$10~meV.

Figures~\ref{fig:compare_AF2_yy}(e) and (f) show the difference between our best model and the data. Our model clearly misses a peak at 3~meV present in the data at both $\QQQ$ positions as well as the 13~meV peak in the data at $\QQQ=(\frac{3}{2},0,\frac{1}{2})-{\bm \epsilon}$ discussed above. The simulations do not include the magnetic order Bragg diffraction so the elastic peak observed in Fig.~\ref{fig:compare_AF2_yy}(d) is not reproduced. 

\section{Discussion}

\subsection{Magnetic Hamiltonian}
 Our model and neutron scattering data in the AF1 phase are in very good agreement. We find, as expected, that by far the largest interaction is the antiferromagnetic exchange along the chains, which leads to a fractionalisation of the magnons into spinons at high energies, characteristic of the $S=\frac{1}{2}$ Heisenberg AFM chain.

Most of the smaller exchange interactions that have been considered previously are also required in our Hamiltonian in order to describe the observed magnon dispersion in linear spin-wave theory. In particular, the next-nearest-neighbour exchange parameter $J_{2,ab}$ is found to be necessary in order to reproduce the minimum in the dispersion at the X-point. The other next-nearest neighbour exchange interactions, $J_{2,a}$ and $J_{2,c}$, however, are not required to describe our data even though \emph{ab initio} calculations predict that at least one of these interactions should be significant (see Table~\ref{tab:model_parameters}). 

Similarly, several studies indicate that the Dzyaloshinskii--Moriya (DM) interaction should play a prominent role in CuO, particularly in the AF2 phase \cite{Pradipto2012,Jin2012,Giovannetti2011a}. We investigated the effect of the symmetry-allowed DM interactions in the $J$ and $J_{bc}$ bonds using our model and assuming $C2/c$ symmetry. We found that the DM interactions do affect the spin wave energies, as expected. In particular, with $\delta=0$ a DM interaction in the $J$ bond with a magnitude of around 15\,meV can produce a splitting of the optic and acoustic modes of similar magnitude to that observed experimentally in the AF1 phase. However, the calculated mode intensities did not even qualitatively match the polarized neutron scattering data presented in Fig.~\ref{fig:neutron_Iyy_lowT}. We conclude, therefore, that the DM interaction is not responsible for the splitting of the optic and acoustic modes. Smaller values of the DM interaction had less impact on the model, but we were unable to find any magnitude or direction of this interaction that improved the fit. We also attempted to improve the model of the AF2 phase  by inclusion of a DM interaction but were not successful. Although our findings do not imply that interactions such as $J_{2,a}$, $J_{2,c}$ and the DM interaction are negligible, they do indicate that interactions neglected in our model are sufficiently small not to impact on the magnon spectrum to within the sensitivity of our measurements.

Our results confirm the importance of magnetic frustration in CuO. The AF1 phase is composed of two AFM lattices with strongly frustrated inter-lattice interactions. This frustration is partially relieved in the AF1 phase, splitting the magnetic modes into acoustic and optic modes. The nature of the interaction which relieves the frustration and stabilises the AF1 phase is not determined here conclusively, but we have been able to successfully reproduce the observed optic--acoustic mode splitting by assuming that the nearest-neighbor inter-lattice interaction $J_{bc}$ splits into two unequal interactions $J_{bc}^-$ and $J_{bc}^+$ which differ by $\delta$, see Fig.~\ref{fig:J_bc_split} and Eq.~(\ref{eq:delta}).  We note that the inter-lattice coupling could equally be obtained by splitting the $J_{ab}$ and $J_{ab}'$ interactions (see Fig.~\ref{fig:exchange_interactions}(b)). The choice to split $J_{bc}$ rather than $J_{ab}$ was arbitrary.

The introduction of $\delta$ (or an equivalent parameter for $J_{ab}$/$J_{ab}'$) reduces the symmetry of the lattice. Physically, this broken symmetry could arise from an as-yet undetected structural distortion in the AF1 phase. This symmetry-breaking is consistent with the doubling of the unit cell $\{ \aaa,\bbb,\ccc \} \rightarrow \{ \aaa+\ccc,\bbb,\aaa-\ccc \}$ suggested by infrared spectroscopy measurements\cite{Kuzmenko2001,Choi2013}, which is itself identical to the magnetic unit cell in the AF1 phase, shown in Fig~\ref{fig:exchange_interactions}. The highest symmetry space group consistent with our results is $P21/c$. It would be interesting to perform high resolution diffraction measurements to search for such a structural distortion.

In order to model the spectra at higher temperatures we find it necessary to reduce $\delta$. This implies an increase in the inter-lattice frustration with increasing temperature.
In the AF2 phase the effect of a small but non-zero $\delta$ is to cause alternate spins along the chains of the helix to tilt by an angle $\phi$, see Figs.~\ref{fig:helix_illustration}(c) and (d). More detailed diffraction measurements than currently exist would be needed to detect this small deviation from the reported AF2 spin structure.

The agreement between our model and the AF2 phase neutron data is not very satisfactory, see Figs.~\ref{fig:compare_AF2_yy}(c)--(f). We find additional peaks in the data at 3~meV and 13~meV that the model does not account for, the latter being observed only at $\QQQ=(\frac{3}{2},0,\frac{1}{2})-{\bm \epsilon}$. Interestingly, the peak at 3~meV coincides with the energy of an electromagnon observed in CuO by THz spectroscopy\cite{Jones2014}. The 3~meV peak in our data could, therefore, be due to scattering from the magnetic component of the electromagnon. Inclusion of spin--lattice coupling in our model would be required to test this possibility.  An electromagnon has also been predicted at 13.5~meV \cite{Cao2015a}, which might explain the feature in our data at 13~meV. Up to now, however, this electromagnon has not been observed by other techniques, and from our modelling we cannot exclude the possibility that our 13~meV feature is not the onset of the optic magnon branch.
  
In the spin-wave theory used here we have not included higher order terms involving three or more spins. 
Such terms have previously been shown to provide plausible explanations of the stability of the AF1 phase and other features of CuO, e.g.~Refs.~\onlinecite{Yablonskii1990,Pasrija2013,Jones2014,Moser2015}. The order-by-disorder mechanism would also favor the collinear AF1 phase\cite{Villain1980,Henley1987,Henley1989}.
It would be interesting to calculate the magnon spectrum from these models to see whether any of the proposed higher-order interactions could provide an alternative mechanism for inter-lattice coupling and the consequent lifting of the degeneracy of the acoustic and optic modes, which in our model is achieved by breaking the symmetry of the frustrated $J_{bc}$ interactions.

\subsection{Magnetic anisotropies}
Our results show that the magnetic anisotropy is of the easy-plane type, the easy plane being the plane in which the spins rotate in the helicoidal AF2 phase. This anisotropy is consistent with the observed field-induced magnetic phases\cite{Wang2016}, but is obtained here directly from the low energy polarized neutron data at low temperature presented in Figs.~\ref{fig:neutron_Iyy_lowT}(a) and (b), which shows that the gap for spin fluctuations parallel to the easy plane is much smaller than that for spin fluctuations perpendicular to the easy plane.

Within the easy plane there is a small axial anisotropy along the $b$ axis. The origin of this anisotropy can be understood if there were a weak spin-orbit-induced single-ion anisotropy which favored the direction normal to the CuO$_4$ plaquettes. The plaquette normals alternate along the chains between $+\theta$ and $-\theta$ to the $b$ axis ($\theta=39^\circ$), while the strong AFM exchange favors collinear alignment of spins. With this type of anisotropy, therefore, energy is lowest when the spins are aligned along the $\pm b$ axis.

A spin-flop transition has been reported at $B=10.4$~T at low temperature for magnetic fields applied along the $b$ axis\cite{Kondo1988,Monod1998,Wang2016}. The spin-flop field $B_\textrm{sf}$ depends on the exchange interactions and the axial anisotropy. In a simple two-spin model with exchange ($J$) and axial anisotropy $|D_1| \ll J$, the spin-flop transition occurs at
\begin{align}
g \mu_B B_\textrm{sf} = 2S \sqrt{|D_1|J}.
\end{align}
Using our values of $D_1=-0.015$~meV and $J=91.4$~meV, we find $B_\textrm{sf} \approx 10$~T, in excellent agreement with the results of  Refs.~\onlinecite{Kondo1988,Monod1998}.

\section{Conclusions}
We have investigated the spin excitation spectrum of CuO over a large volume of reciprocal space. 
Measurements at high energies agree well with the spinon spectrum for a spin-$\frac{1}{2}$ Heisenberg antiferromagnetic chain. The coupling between neighboring chains in CuO is strongly frustrated, but the observation of an optic magnon mode indicates a partial relief of frustration which implies a lowering of symmetry in the magnetically ordered phases,

In directions perpendicular to the chains the spectrum shows well-defined spin-wave excitations which we have successfully modeled using linear spin-wave theory. The results have enabled us to refine an effective spin Hamiltonian for CuO, revealing significant discrepancies with previous models and \emph{ab initio} calculations. The spin Hamiltonian can be used to understand and predict in detail the magnetic properties of CuO, and it forms a platform on which models for the magnetoelectric behaviour can be developed. 

\begin{acknowledgments}
We would like to thank T. Ziman, R. Coldea, X. Rocquefelte and W. Lafargue-dit-Hauret for helpful discussions, and H. M. R{\o}nnow for the loan of a sample mount.  This work was supported by the U.K. Engineering and Physical Sciences Research Council (Grant Nos. EP/J012912/1 and EP/N034872/1, and a studentship for S.M.G.). Experiments at the ISIS Neutron and Muon Source were supported by a beamtime allocation from the Science and Technology Facilities Council.
\end{acknowledgments}

\appendix

\section{Polarised neutron scattering experiments} \label{sec:neutron_detail}
Longitudinal polarisation analysis can be used to measure the following six neutron cross sections (referred to the Blume--Maleev coordinate system):

\begin{align}
 S^x_\text{SF} &= \frac{2}{3}S^\text{sp}_\text{inc}+S^M_{yy}+S^M_{zz} \pm S^M_{yz},\label{eq:IxSF}  \\
 S^y_\text{SF} &= \frac{2}{3}S^\text{sp}_\text{inc}+S^M_{zz}, \\
 S^z_\text{SF} &= \frac{2}{3}S^\text{sp}_\text{inc}+S^M_{yy}, \label{eq:IzSF} \\
 S^x_\text{NSF} &= S^\text{N}_\text{coh}+S^\text{iso}_\text{inc}+\frac{1}{3}S^\text{sp}_\text{inc},\\
 S^y_\text{NSF} &= S^\text{N}_\text{coh}+S^\text{iso}_\text{inc}+\frac{1}{3}S^\text{sp}_\text{inc} \pm S^\text{NM}_y + S^M_{yy},\\
 S^z_\text{NSF} &= S^\text{N}_\text{coh}+S^\text{iso}_\text{inc}+\frac{1}{3}S^\text{sp}_\text{inc}\pm S^\text{NM}_z + S^M_{zz},
\end{align}
where $S^x_\text{SF}$ stands for spin-flip scattering with neutron spin quantization direction parallel to $x$. The other terms are: $S^\text{sp}_\text{inc}$ is spin incoherent scattering, $S^\text{iso}_\text{inc}$ is isotopic incoherent scattering, $S^\text{N}_\text{coh}$ is nuclear coherent scattering, $S^\text{NM}$ is an interference term between nuclear and magnetic scattering. The terms $S^M_{yy}$, $S^M_{zz}$ and $S^M_{yz}$ represent magnetic scattering:
\begin{align}
 S^M_{\alpha \alpha}(\QQQ,\omega) & = \frac{1}{2\pi\hbar} \int_{-\infty}^\infty \langle M_\alpha^\dagger(\QQQ) M_\alpha(\QQQ,t) \rangle\, \textrm{e}^{-\textrm{i} \omega t}\,\mathrm{d}t \\
 S^M_{yz}(\QQQ,\omega) & = \frac{1}{2\pi\hbar} \int_{-\infty}^\infty \langle M_y^\dagger(\QQQ) M_z(\QQQ,t) \nonumber\\
 & \ \ \ \ \ \ \ \ \ \ \ \ \ \ \ \ -  M_z^\dagger(\QQQ) M_y(\QQQ,t)  \rangle\, \textrm{e}^{-\textrm{i} \omega t}\,\mathrm{d}t,
\end{align}
where $\alpha=y,z$ and $M_\alpha(\QQQ,t)$ is the Fourier transform of the $\alpha$ component of the magnetization. The $\pm$ sign before $S_{yz}^M$ in Eq.~\eqref{eq:IxSF} refers to the direction of the incoming polarisation relative to the quantization axis.

From these equations we see that
\begin{align}
 S^x_\text{SF}-S^y_\text{SF} &= S^M_{yy} \pm S^M_{yz}, \label{eq:Ix-Iy}\\
 S^x_\text{SF}-S^z_\text{SF} &= S^M_{zz} \pm S^M_{yz}. \label{eq:Ix-Iz}
\end{align}

In some cases the $ S^x_\text{SF}$ cross section was not measured, and thus $S^M_{yy}$ was estimated using Eq.~(\ref{eq:IzSF}) and subtracting a constant from the data for the $S^\text{sp}_\text{inc}$ term.
The flipping ratios for all three polarisation directions measured with $k_\textrm{f} = 2.662$~\AA{}$^{-1}$ on the $20\bar{2}$ nuclear Bragg reflection were $R_x = 12$, $R_y = 12$ and $R_z = 21$, corresponding to polarisation efficiencies of  85--91\%.

There were no distinctive features in the non-magnetic data except for a phonon at around 15~meV. The $S^M_{yz}$ term is zero in the AF1 phase. In the AF2 phase, the sign of this term depends on the chirality of the helix. A macroscopic sample such as ours will have near equal populations of domains with positive and negative chirality, and hence the $S^M_{yz}$ term will not be measurable in the AF2 phase. 

\section{Additional neutron scattering data and fits}
\label{sec:neutron_exp_data}

\begin{figure}[t]
\includegraphics[width=0.42\textwidth]{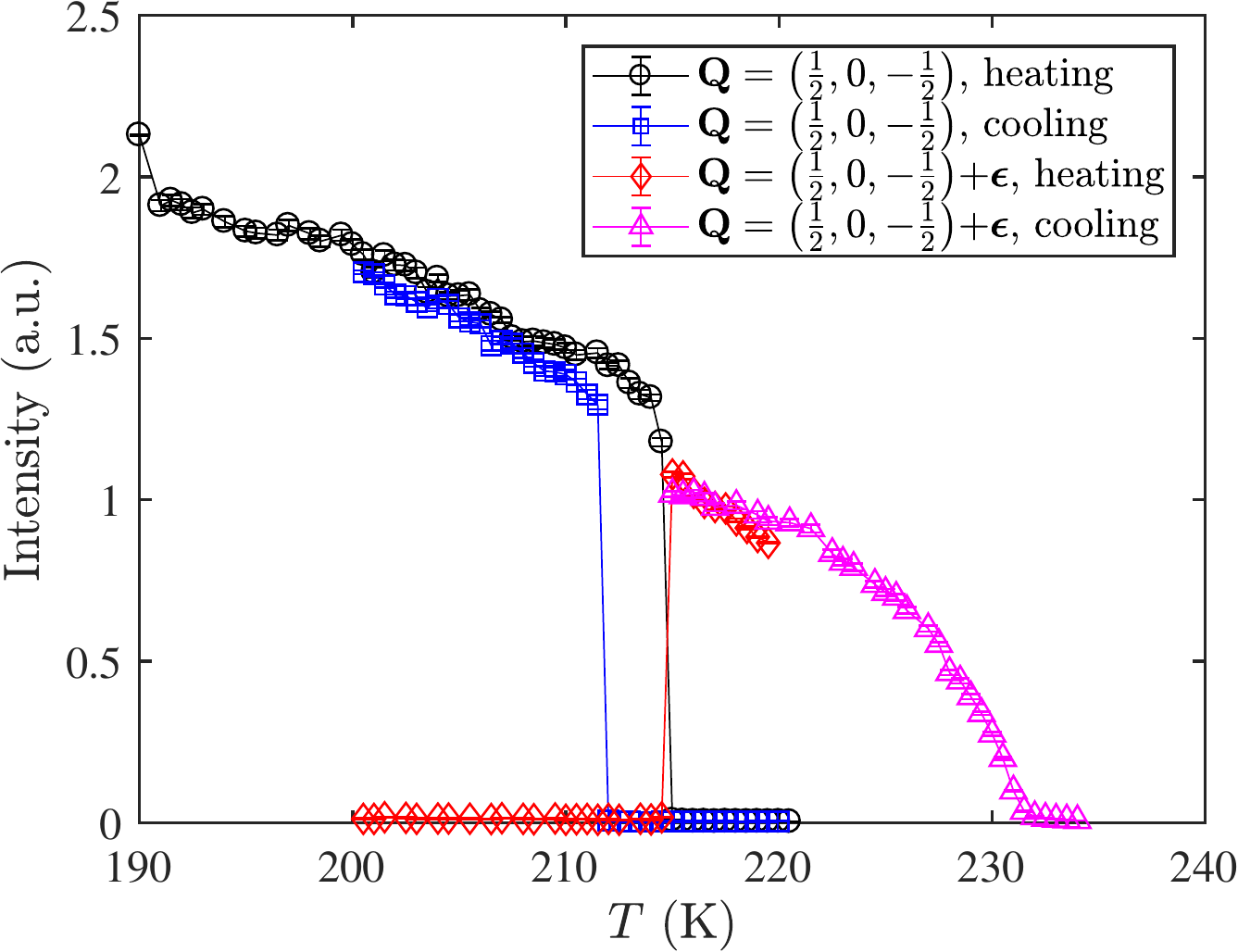}
\caption{The temperature dependence of $S_\textrm{SF}^x$, Eq.~\eqref{eq:IxSF}, measured at magnetic Bragg peaks in the AF1 and AF2 phases. The first order transition between the AF1 and AF2 phases is clearly seen. Data at $\QQQ=(\frac{1}{2},0,-\frac{1}{2})+{\bm \epsilon}$ were not measured below 215 K upon cooling.
}
\label{fig:temperature_dependence}
\end{figure}

\begin{figure}[t]
 \centering
  \includegraphics[width=0.42\textwidth]{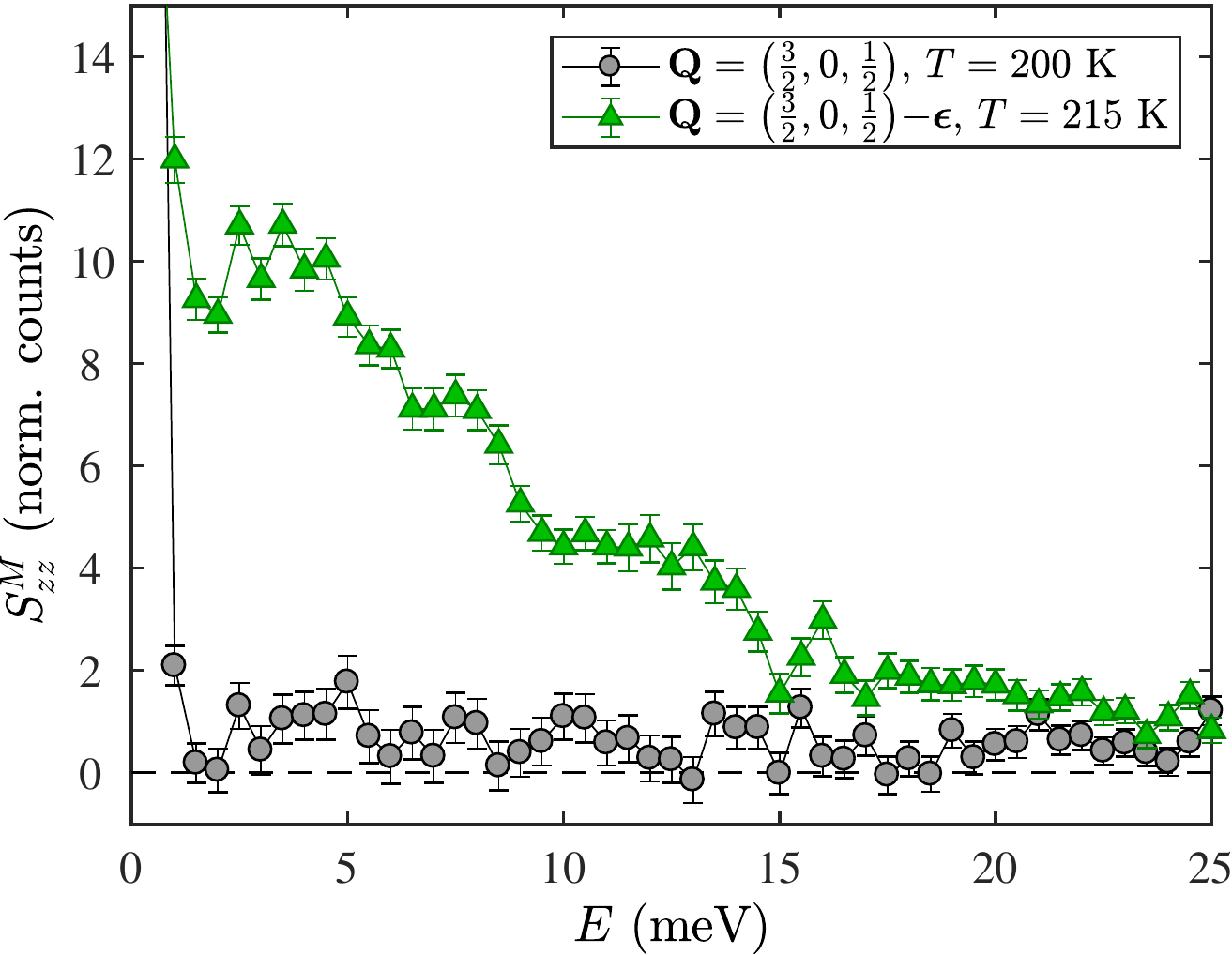}
 \caption{Magnetic response function   $S^M_{zz}$ of CuO measured at elevated temperatures by neutron polarisation analysis near $\textbf{Q} = (\frac{3}{2},0,-\frac{1}{2})$. The symbols show   $S^M_{zz}$ deduced from the $x$ and $z$ spin-flip (SF) scattering, Eq.~(\ref{eq:Ix-Iz}). }
 \label{fig:compare_AF2_zz}
\end{figure}

The temperature dependence of the magnetic order Bragg peaks is shown in Fig.~\ref{fig:temperature_dependence}. The discontinuous transition between the AF1 and AF2 phases is clearly displayed. There is a hysteresis of about 3\,K which may partly reflect the first order nature of the transition but is most likely dominated by the lag between the sample temperature and sensor during heating/cooling. The CuO crystal was large and is an insulator, so although the crystal was in helium exchange gas the time to reach thermal equilibrium would have been long compared with the time scale of the measurements during the temperature sweeps.

In Fig.~\ref{fig:compare_AF2_zz} we show the magnetic response function  $S^M_{zz}$ measured at 200~K in the AF1 phase and at 215~K in the AF2 phase. In the AF1 phase there is essentially no inelastic signal, consistent with the magnetic moments ordered along the $b$ axis, which is parallel to $z$. Conversely, in the AF2 phase the $S^M_{zz}$ signal resembles that observed in the $S^M_{yy}$ channel, Fig.~\ref{fig:compare_AF2_yy}(d).

\begin{figure*}[t]
 \centering
 \includegraphics[width=0.81\textwidth]{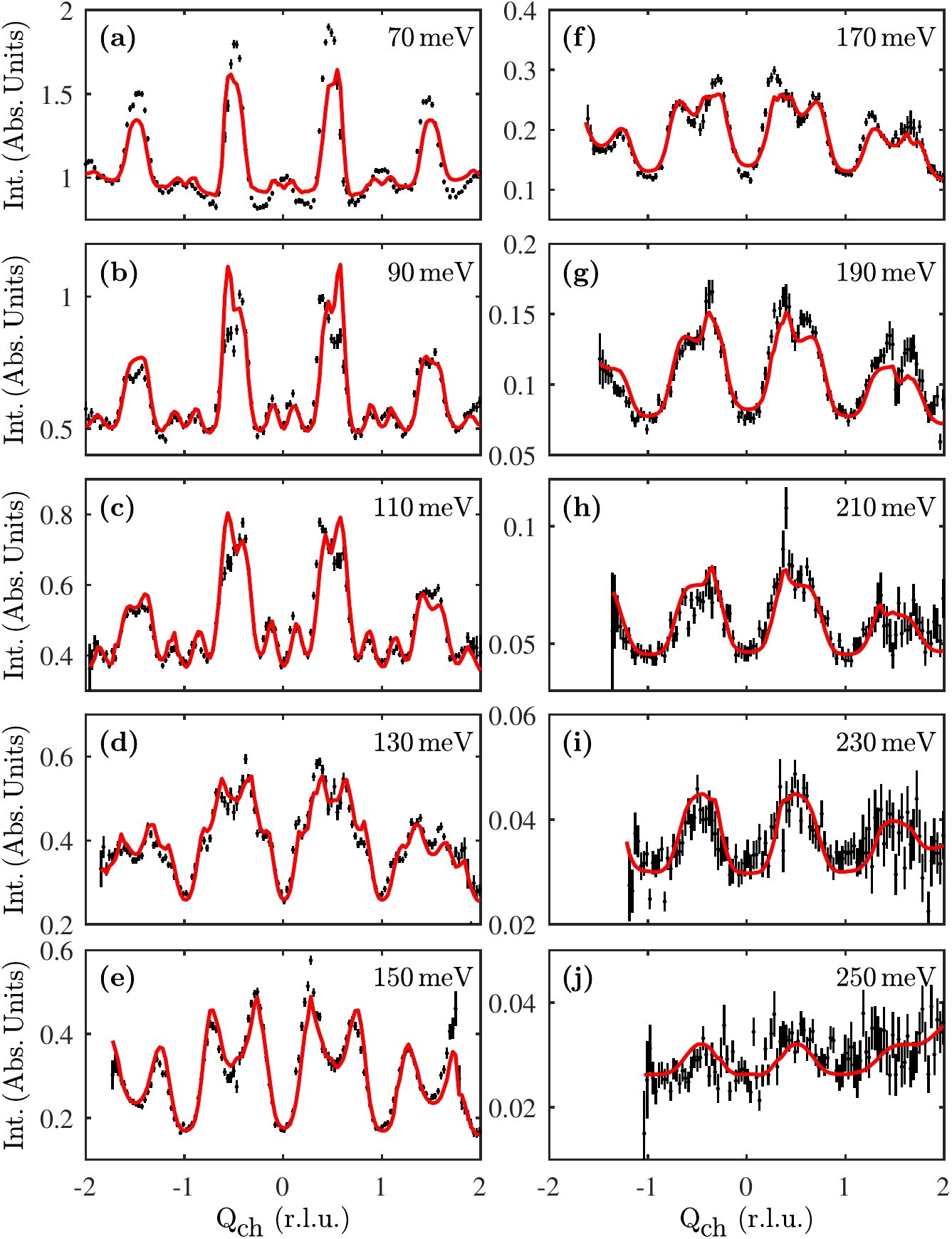}
 \caption{Constant-energy cuts across the spinon continuum spectrum of CuO. Data points (black markers) were measured on the MAPS spectrometer using $E_i = 300$~meV. The line cuts are measured at constant energies increasing from (a) to (j). All cuts are integrated across a 20\,meV range to
improve statistics. The average energy $E$ after this integration is recorded in each sub-panel. All intensities are recorded in absolute units of mb sr $^{-1}$ meV $^{-1}$ f.u. $^{-1}$ . The global fit to the M\"u{}ller ansatz convolved with the MAPS resolution function is shown
by the red lines.}
\label{fig:spinon_1d_fits}
\end{figure*}

In Fig.~\ref{fig:spinon_1d_fits} we show multiple constant-energy  cuts across the high energy part of the spectrum of CuO together with our fit to the M\"u{}ller ansatz convolved with the MAPS resolution function. In general, the agreement between the data and the model is excellent. There are some discrepancies, most prominently at the lowest energy shown, 70~meV. We find that these discrepancies coincide with the boundaries of the detector and can be ascribed to systematic errors\cite{GawThesis}.

\section{Analytic spin-wave calculations} 
\label{sec:analytical}

We give an analytic expression for the spin-wave dispersion in the AF1 phase in CuO. The calculations are performed for the Heisenberg Hamiltonian Eq.~(\ref{eq:Heisenberg_simple}) together with easy-axis anisotropy, but do not include the easy-plane anisotropy. We included all the relevant exchange constants. This model gives a very good description of most of the spin-wave spectrum in the AF1 phase, with the exception of the anisotropic signal at low energy caused by the easy-plane anisotropy which splits the pairs of modes. Details of the calculations are presented in Ref.~\onlinecite{GawThesis}, and the model is similar to that in Ref.~\onlinecite{Wheeler2009}.\\

Diagonalization of the Hamiltonian by the standard Holstein--Primakov 
method for the eight-spin magnetic cell shown in Fig.~\ref{fig:exchange_interactions}(a) gives two fourfold-degenerate modes with dispersion relations given by
\begin{widetext}
\begin{align}
E_\pm^2=A^2 + |B|^2 -C^2 -|D|^2 \pm \sqrt{4|AB-CD^*|^2-|B^*D^* - BD|^2},
\end{align}
where 
\begin{align}
A&=2S[J+J_{2,a}+J_{2,c}+2J_{2,ab}+J^+_{bc}-J^-_{bc}+J_{ac}(\cos(\pi(H+L))-1)+J_b(\cos(2\pi K)-1)     -D_1], \nonumber\\
B&=S[J_{ab}(e^{i\pi(H+K)}+e^{i\pi(H-K)})+J^-_{bc}(e^{-i\pi(L-K)}+e^{-i\pi(L+K)})],\nonumber\\
C&=2S[J\cos(\pi(H-L))+J_{2,c}\cos(2\pi L)+J_{2,a}\cos(2\pi H) +J_{2,ab} \{\cos(2\pi (H+L))+\cos(2\pi (H-L)) \}  ],\nonumber\\
D&=S[J_{ab}(e^{i\pi(H+K)}+e^{i\pi(H-K)})+J^+_{bc}(e^{-i\pi(L-K)}+e^{-i\pi(L+K)})],\nonumber
\end{align}
\end{widetext}
and the scattering vector $\QQQ=(H,K,L)$ is given in reciprocal lattice units.

\section{Mean-field model of the AF2 phase} \label{sec:helix_calculation}
In a chain in which the next-nearest-neighbor interaction is strongly antiferromagnetic and the nearest-neighbor interaction is weak, a helical arrangement of the spins is preferred. The pitch of the helix depends on the relative size of the exchange interactions.

In the AF2 phase, CuO can be seen as consisting of such chains, where $J$ is the next-nearest neighbor interaction and $J_{ab}$ and $J_{bc}$ are nearest-neighbor interactions. $J_{ac}$ and $J_{2,ab}$ couple neighboring chains on the same lattice.

We let $n$ label spins along the $a$ direction, and $m$ label spins along the $c$ direction. $\qqq=(q_a,0,q_c) = (0.506,0,-0.483)$ is the propagation vector in reciprocal lattice units. The propagation vector has no component along the $b$ axis, which allows us to omit this coordinate for simplicity. 
We neglect the small axial anisotropy within the plane of rotation of the spins in these calculations. The easy-plane anisotropy defines the coordinate system so that $x$ and $y$ are in the plane of the helix. There is no out of plane component of the spins and the easy plane thus merely adds a constant to the energy, which we ignore here.

The Hamiltonian is then \begin{align}
\mathcal{H} &= \sum_{n,m}J\SSS_{n,m}\cdot \SSS_{n+1,m-1}  + 2J_{2,ab} \SSS_{n,m} \cdot \SSS_{n+2,m} \nonumber \\
&+J_{bc}^+ \SSS_{n,m} \cdot \SSS_{n,m-1}+ J_{bc}^- \SSS_{n,m} \cdot \SSS_{n,m+1} \nonumber \\
&+J_{ac}\SSS_{n,m}\SSS_{n+1,m+1} + 2J_{ab} \SSS_{n,m} \cdot \SSS_{n+1,m}.
\end{align}
The factor of 2 for the $J_{ab}$ and $J_{2,ab}$ terms is to take into account the interaction along the $[1,1,0]$ and $[1,-1,0]$ directions.
The splitting of $J_{bc}$ changes the total effective near-neighbor interaction in neighboring chains, which therefore will have a phase difference. We therefore write 



\begin{align}
S^x_{n,m} &= S\cos(n\pi q_a  + m \pi q_c ) \text{ for } n+m \text{ even}\\
S^y_{n,m} &= S\sin(n\pi q_a  + m \pi q_c )  \text{ for } n+m \text{ even}, \\
S^x_{n,m} &= S\cos(n\pi q_a  + m \pi q_c +\phi) \text{ for } n+m \text{ odd}, \\
S^y_{n,m} &= S\sin(n\pi q_a  + m \pi q_c +\phi)  \text{ for } n+m \text{ odd},
\end{align}
where $a$ and $c$ are the lattice constants. 

The mean field energy per spin pair,  $E_2$, is
\begin{align}
E_2/S^2 &= J\cos(2\pi q_a - 2\pi q_c )  +J_{ac}\cos(2\pi q_a + 2\pi q_c )\nonumber \\
&+ J_{ab} (\cos(2\pi q_a+\phi) + \cos(2\pi q_a-\phi))\nonumber \\
&+J_{bc}^+\cos(2\pi q_c +\phi) + J_{bc}^-\cos(2\pi q_c -\phi) \nonumber \\
&+ 2J_{2,ab} \cos(4\pi q_a).
\end{align}

The AF2 phase is most stable when this expression is at a minimum. Minimizing with respect to $2\pi q_a$, $2\pi q_c $ and $\phi$ gives three equations to determine $J_{ab}$, $J_{bc}$ and $\phi$ ($\delta$ is determined from the neutron scattering data).

The equations are
\begin{align}
\frac{d E_2}{d 2\pi q_a} &= -J\sin(2\pi q_a - 2\pi q_c ) -J_{ac}\sin(2\pi q_a + 2\pi q_c )  \label{eq:dE_dqa} \\
&-4J_{2,ab}\sin(4\pi q_a)- 2J_{ab} \sin(2\pi q_a) \cos(\phi)=0,\nonumber
\end{align}
\begin{align}
\frac{d E_2}{d 2\pi q_c } &= J\sin(2\pi q_a - 2\pi q_c ) -J_{ac}\sin(2\pi q_a + 2\pi q_c )  \label{eq:dE_dqc} \\
&- 2J_{bc} \sin(2\pi q_a) \cos(\phi)-\delta\cos(2\pi q_c ) \sin(\phi)=0,\nonumber
\end{align}
and
\begin{align}
\frac{d E_2}{d \phi} &=-2\sin(\phi) (J_{ab} \cos(2\pi q_a) + J_{bc} \cos(2\pi q_c )) \nonumber \\
&-\delta \sin(2 \pi q_c)\cos(\phi)=0  \label{eq:dE_dphi}.
\end{align}

For $\delta=0$ we have $\phi=0$ and these equations are straightforward to solve analytically. For $\delta>0$, numerical methods must be applied. In general, the result is that $J_{bc}<J_{ab}<0$, and the magnitudes of both parameters increase roughly linearly when increasing $\delta$. $\phi$ depends strongly on $\delta$, with $\phi\approx 32^\circ$ for $\delta=0.15$~meV.

The calculations here are consistent with the numerical results from SpinW.

The mean field energy of the AF1 phase is
\begin{align}
E_1/S^2=-J-2J_{2,ab}+J_{ac} -\delta.
\end{align}

The energy of the AF2 phase is lower than that of the AF1 phase for any value of $\delta$ when Eqs.~\eqref{eq:dE_dqa}, \eqref{eq:dE_dqc},\eqref{eq:dE_dphi} are satisfied. On the other hand, if the exchange parameters are kept fixed to their values that are consistent with $\delta=0.15$ as found experimentally at 215~K, and we allow $\delta$, $\phi$ and $\qqq$ to vary, the AF1 phase has lower energy than the AF2 phase when $\delta \gtrsim 0.28$~meV.

\bibliography{cuo,CuO_data}

\end{document}